\documentclass[aps,preprint,nofootinbib,superscriptaddress]{revtex4}%
\usepackage{amssymb}
\usepackage{amsmath}
\usepackage{epsfig}
\usepackage{amsfonts}
\usepackage{graphicx}
\usepackage{hyperref}%
\providecommand{\U}[1]{\protect\rule{.1in}{.1in}}
\begin{document}
\title{Overview on High energy String Scattering Amplitudes and Symmetries of String Theory}
\author{Jen-Chi Lee}
\email{jcclee@cc.nctu.edu.tw}
\affiliation{Department of Electrophysics, National Chiao-Tung University, Hsinchu, Taiwan, R.O.C.}
\author{Yi Yang}
\email{yiyang@mail.nctu.edu.tw}
\affiliation{Department of Electrophysics, National Chiao-Tung University, Hsinchu, Taiwan, R.O.C.}
\date{\today }

\begin{abstract}
We overview symmetries of string scattering amplitudes in the high energy
limits of both the fixed angle or Gross regime (GR) and the fixed momentum
transfer or Regge regime (RR). We calculated high energy string scattering
amplitudes (SSA) at arbitrary mass levels for both regimes. We discovered the
infinite linear relations among fixed angle string amplitudes and the infinite
recurrence relations among Regge string amplitudes. The linear relations we
obtained in the GR corrected the saddle point calculations by Gross, Gross and
Mende. In addition, for the high energy closed string scatterings, our results
differ from theirs by an oscillating prefactor which was crucial to recover
the KLT relation valid for all energies.

We showed\ that all the high energy string amplitudes can be solved by the
linear or recurrence relations so that all the string amplitudes can be
expressed in terms of a single string amplitude. We further found that, at
each mass level, the ratios among the fixed angle amplitudes can be extracted
from the Regge string scattering amplitudes.

Finally, we review the recent developments on the discovery of infinite number
of recurrence relations valid for \textit{all} energies among Lauricella SSA.
The symmetries or relations among SSA at various limits obtained previously
can be\ exactly reproduced. It leads us to argue that the known $SL(K+3,C)$
dynamical symmetry of the Lauricella function maybe crucial to probe spacetime
symmetry of string theory.


\end{abstract}
\maketitle
\tableofcontents
%

\setcounter{equation}{0}
\renewcommand{\theequation}{\arabic{section}.\arabic{equation}}%

\section{Introduction}

Quantum Field Theory (QFT) is a powerful theory in modern physics. The
microcosmic physics is successfully described by using standard model of
particle physics based on QFT. Various experiments have confirmed all
important predictions by standard model in rather precise level. Nevertheless,
the crucial procedure to solve the UV divergence problem in QFT, i.e.
renormalization, is mathematically complicated and has not been fully
understood. Furthermore, the renormalization procedure does not work for
gravity, so\ that a consistent quantum gravity theory\ is impossible to be
constructed by using the conventional QFT. It is usually believed that the
divergence in QFT is due to the topological structure of point-like particles
that cannot be removed without modifying its topological structure. In string
theory, a point-like particle is extended to a small piece of a string, which
changes the topological structure of the theory. The Feynman diagram\ of
strings interaction becomes a smooth world-sheet instead of the QFT world-line
with singularity at interacting points\textsf{.}

To clarify the UV divergence problem in QFT, let us briefly examines the high
energy behavior\ of a Feynman diagram by a simple power counting. In high
energy hard limit, the tree diagram of particles scattering by interchanging a
spin-$J$ particle behaves as $A_{tree}^{\left(  J\right)  }\sim E^{-2(1-J)}$,
so that a one-loop diagram behaves as%
\begin{equation}
A_{1-loop}^{\left(  J\right)  }\sim\int d^{4}p\frac{\left(  A_{tree}%
^{(J)}\right)  ^{2}}{\left(  p^{2}\right)  ^{2}}\sim\int E^{-4(2-J)}\text{
}d^{4}E,
\end{equation}
which is manifestly finite for scalar particles ($J=0$), renormalizable for
vector particles ($J=1$), but nonrenormalizable for particles with $J\geq2$,
including graviton ($J=2$). Nevertheless, there is a loophole in this simple
argument. Assuming that the interchanging states could have different spins,
we thus should sum them all and the final amplitude becomes%
\begin{equation}
A_{tree}=\sum_{J}A_{tree}^{\left(  J\right)  }\sim\sum_{J}a_{J}E^{-2(1-J)},
\end{equation}
which has a essential singularity as $E\rightarrow\infty$ and could behave
rather soft, so that loop amplitudes would be finite, provided the following
two conditions are satisfied simultaneously \cite{YY}:

\begin{enumerate}
\item there are infinite higher spin $J$ particles

\item the coefficients $a_{J}$'s\ are precisely related to each other.
\end{enumerate}

In string theory, the behavior of the string scattering amplitudes was known
to be very soft exponential fall-off in the fixed-angle high energy limit
comparing to the power law behavior of a local quantum field theory.
Therefore, it is natural for us to believe that string theory satisfies the
above two conditions.

String theory trivially satisfied the first condition because a string has
infinite oscillation modes and each of them corresponds to a state with
different spin. However, the second condition is highly nontrivial. We thus
conjecture that it corresponds to a huge hidden symmetry in string theory.

Indeed, there was an evidence which showed that the huge hidden symmetries
proposed was closely related to the softness of string scattering amplitudes
in the hard scattering limit. In \cite{Compact2,Compact}, for string
scatterings in the compact space, the authors discovered the existence of a
power-law regime at fixed angle and simultaneously the linear relations
(symmetry) break down in this regime.

To determine the form of the interaction in a local quantum field theory, a
symmetry principle was postulated beforehand. While in string theory, on the
contrary, it was the interaction, which was prescribed by the very tight
quantum consistency conditions due to the extendedness of one dimensional
string, which determines the form of the symmetry.

It is well known that, in local gauge field theories, symmetries are
spontaneously broken at low energy, but are restored at high energies.
Motivated by this high energy behavior, historically, the first key progress
to uncover symmetries of string theory was to study the high energy, fixed
angle behavior of the hard string scattering (HSS) amplitudes \cite{GM,GM1,
Gross, Gross1,GrossManes} instead of the low energy field theory ones. In the
pioneer work of Gross in 1988 \cite{Gross,Gross1},he made two main conjectures
on this subject. The first one was that at high energies, fixed angle regime
or Gross regime (GR) of the theory, there existed an infinite number of linear
relations among the string scattering amplitudes of different string states at
each order in string perturbation theory \ The other one was that the
scattering amplitudes of all the infinite number of string states can be
determined in terms of one single dilaton (tachyon for the case of open
string) scattering amplitudes by this symmetry. Nevertheless, the symmetry
charges of his proposed high energy stringy symmetries were not understood and
the ratios among scattering amplitudes of different string states were not calculated.

The second key progress to uncover symmetries of string theory was the
realization of the importance of zero norm states (ZNS) in the old covariant
first quantized (OCFQ) string spectrum. In the works of
\cite{Lee,Lee-Ov,LeePRL}, it was proposed that spacetime symmetry charges of
string theory were originated from an infinite number of ZNS with arbitrary
high spins in the old covariant quantized string spectrum.

This review is organized as following. In chapter II, we discuss and calculate
stringy symmetries which were calculated to be valid for all energies. These
calculations include stringy symmetries calculated by (1) $\sigma$-model
approach of string theory in the first order weak field approximation, (2)
Discrete ZNS and $w_{\infty}$ symmetry of $2D$ string theory and (3) Soliton
ZNS and the corresponding enhanced stringy gauge symmetries. We will
concentrate on using the idea of ZNS and its applications to various
calculations of stringy symmetries.

In chapter III, we will calculate high energy, fixed angle HSS amplitudes. The
hard stringy Ward identities derived from the decoupling of ZNS in the HSS
limit will be used to explicitly prove Gross's two conjectures
\cite{ChanLee,ChanLee1,ChanLee2, CHL, CHLTY1,CHLTY2}. An infinite number of
linear relations among hard string scattering amplitudes of different string
states were then derived. Remarkably, these linear relations can be used to
determine the proportionality constants or ratios among HSS amplitudes of
different string states algebraically at each fixed mass level.

In chapter IV, we discuss the hard closed string scatterings. We will point
out and clarify the inconsistency of the calculation of Gross and Mende
\cite{GM,GM1} by using the KLT relation \cite{KLT} which is valid for all
energies. The first "string BCJ relation" was then discovered \cite{Closed} in
2006, and was independent of the discovery of field theory BCJ relation
\cite{BCJ1} in 2008. We will also discuss hard string scattered from
D-branes/O-planes and in compact space, and study their high energy behaviors.

In chapter V, we will calculate another high energy string scattering
amplitudes, the Regge string scattering (RSS) amplitudes. We will see that, in
contrast to the linear relations in the GR, there exists an infinite number of
"recurrence relations" among RSS amplitudes of different string states. These
recurrence relations can be used to solve all RSS amplitudes and express them
in terms of one single amplitude. Moreover, an interesting link between the
HSS and the RSS amplitudes was discovered, and the ratios among fixed angle
amplitudes can be extracted from RSS amplitudes \cite{bosonic,RRsusy}.

In chapter VI, we briefly review the recent developments on the discovery of
an infinite number of recurrence relations which were valid for \textit{all}
energies among the \textit{exact} open bosonic string scattering amplitudes of
three tachyons and one arbitrary string state, or the so-called Lauricella
string scattering amplitudes (LSSA) \cite{LLY2}. These infinite number of
recurrence relations can be used to solve \cite{LLLY} all the LSSA
algebraically and express them in terms of one single four tachyon amplitude.

Moreover, string scattering amplitudes and symmetries or relations among
string scattering amplitudes among different string states at various
scattering limits calculated previously can be rederived. These include the
stringy linear relations conjectured by Gross and proved by Taiwan group
\cite{ChanLee,ChanLee1,ChanLee2, CHL, CHLTY1,CHLTY2} in the hard scattering
limit, the recurrence relations in the Regge scattering limit
\cite{AppellLY,LY} and the extended recurrence relations in the
nonrelativistic scattering limit discovered recently \cite{LLY1}. Finally, a
conclusion is given in chapter VII.

For more detailed review on high energy string scattering amplitudes and
symmetries of string theory, see the recent long review paper \cite{LYreview}.

\section{Zero norm states and enlarged stringy symmetries}

In the calculation of $\sigma$-model approach of string theory, one turns on
background fields on the worldsheet energy momentum tensor $T$. The conformal
invariance of the worldsheet then requires, in addition to $D=26$, the
cancellation of various kinds of $q$-number anomalies and results to equations
of motion of the background fields \cite{GSW}. It was then demonstrated that
\cite{Lee} for each \textit{spacetime} ZNS, one can systematically construct a
\textit{worldsheet} $(1,1)$ primary field $\delta T_{\Phi}$ such that%
\begin{equation}
T_{\Phi}+\delta T_{\Phi}=T_{\Phi+\delta\Phi} \label{BZNS}%
\end{equation}
is satisfied to some order of weak field approximation in the $\sigma$-model
background fields $\beta$ function calculation. In the above equation,
$T_{\Phi}$ is the worldsheet energy momentum tensor with background fields
$\Phi$ and $T_{\Phi+\delta\Phi}$ is the new energy momentum tensor with the
new background fields $\Phi+\delta\Phi$. As a result, for each ZNS one can
construct a spacetime symmetry transformation for the corresponding string
background fields.

In addition to the positive norm physical propagating modes, there are two
types of physical ZNS in the old covariant first quantized (OCFQ) open bosonic
string spectrum: \cite{GSW}%
\begin{align}
\text{Type I}  &  :L_{-1}\left\vert x\right\rangle ,\text{ where }%
L_{1}\left\vert x\right\rangle =L_{2}\left\vert x\right\rangle =0,\text{
}L_{0}\left\vert x\right\rangle =0;\\
\text{Type II}  &  :\left(  L_{-2}+\frac{3}{2}L_{-1}^{2}\right)  \left\vert
\widetilde{x}\right\rangle ,\text{ where }L_{1}\left\vert \widetilde{x}%
\right\rangle =L_{2}\left\vert \widetilde{x}\right\rangle =0,\text{ }%
(L_{0}+1)\left\vert \widetilde{x}\right\rangle =0.
\end{align}
While type I states have zero-norm at any spacetime dimension, type II states
have zero-norm \emph{only} at $D=26$. For example, among other stringy
symmetries, an \textit{inter-particle} symmetry transformation for two
propagating states at mass level $M^{2}=4$ of open bosonic string can be
generated \cite{Lee}
\begin{equation}
\delta C_{(\mu\nu\lambda)}=\frac{1}{2}\partial_{(\mu}\partial_{\nu}%
\theta_{\lambda)}^{2}-2\eta_{(\mu\nu}\theta_{\lambda)}^{2},\delta C_{[\mu\nu
]}=9\partial_{\lbrack\mu}\theta_{\nu]}^{2}, \label{01}%
\end{equation}
where $\partial^{\mu}\theta_{\mu}^{2}=0,(\partial^{2}-4)\theta_{\mu}^{2}=0$
which are the on-shell conditions of the $D_{2}$ vector ZNS with polarization
$\theta_{\mu}^{2}$ \cite{Lee}%

\begin{equation}
|D_{2}\rangle=[(\frac{1}{2}k_{\mu}k_{\nu}\theta_{\lambda}^{2}+2\eta_{\mu\nu
}\theta_{\lambda}^{2})\alpha_{-1}^{\mu}\alpha_{-1}^{\nu}\alpha_{-1}^{\lambda
}+9k_{\mu}\theta_{\nu}^{2}\alpha_{-2}^{[\mu}\alpha_{-1}^{\nu]}-6\theta_{\mu
}^{2}\alpha_{-3}^{\mu}]\left\vert 0,k\right\rangle ,\text{ \ }k\cdot\theta
^{2}=0, \label{02}%
\end{equation}
and $C_{(\mu\nu\lambda)}$ and $C_{[\mu\nu]}$ are the two background fields of
the symmetric spin-three and antisymmetric spin-two propagating states respectively.

In the higher mass levels, $M^{2}=6$ for example, a new phenomenon begins to
show up. There are ambiguities in defining positive-norm spin-two and scalar
states due to the existence of ZNS in the same Young representations
\cite{LeePRL}. As a result, the degenerate spin two and scalar positive-norm
states can be gauged to the higher rank fields, the symmetric spin four
$D_{\mu\nu\alpha\beta}$ and mixed-symmetric spin three $D_{\mu\nu\alpha}$ in
the first order weak field approximation. In fact, it was shown \cite{Lee3}
that the scattering amplitude involving the positive-norm spin-two state could
be expressed in terms of the of spin-four and mixed-symmetric spin-three
states due to the existence of a \textit{degenerate} type I and a type II
spin-two ZNS. Presumably, this stringy phenomenon seems to persist to higher
mass levels.

This calculation is consistent with the result in the HSS limit. In fact, it
can be shown that in the HSS limit all the scattering amplitudes of leading
order in energy at each fixed mass level can be expressed in terms of those of
the leading trajectory string states with transverse polarizations on the
scattering plane. See the calculations of Eq.(\ref{03}), Eq.(\ref{055}) and
Eq.(\ref{04}) in chapter III. One can also justify this decoupling phenomenon
by WSFT \cite{KaoLee}. Finally one expects this decoupling to persist even if
one includes the higher order corrections in the calculation of weak field
approximation, as there will be even stronger relations among background
fields order by order through iteration.

The calculation of Eq.(\ref{01}) obtained in the first order weak field
approximation is valid for all energies or all orders in $\alpha^{\prime}$.
The second order calculation in the weak field approximation implies an even
more interesting spontaneously broken inter-mass level symmetry in string
theory \cite{Lee4,LEO}. Some implications of the corresponding stringy Ward
identities on the string scattering amplitudes were discussed in
\cite{JCLee,Lee4}. On the other hand, it was then realized that \cite{KaoLee,
CLYang} the symmetry in Eq.(\ref{01}) can be reproduced from the off-shell
gauge transformations of Witten string field theory (WSFT) \cite{Witten} by
imposing the no ghost conditions. It is important to note that this stringy
symmetry exists only for $D=26$ due to the type II ZNS in the OCFQ string
spectrum , which only exists for $D=26$.

Incidentally, it was well known in toy $2D$ string theory that the operator
products of the discrete positive norm states \cite{ChungLee1}%
\begin{equation}
\psi_{J,M}^{\pm}\sim\left\vert
\begin{array}
[c]{cccc}%
S_{2J-1} & S_{2J-2} & \cdots & S_{J+M}\\
S_{2J-2} & S_{2J-3} & \cdots & S_{J+M-1}\\
\vdots & \vdots & \ddots & \vdots\\
S_{J+M} & S_{J+M-1} & \cdots & S_{2M+1}%
\end{array}
\right\vert \cdot\exp\left[  \sqrt{2}(iMX(0)+(-1\pm J)\phi(0))\right]
\label{3.5.}%
\end{equation}
form a $w_{\infty}$ algebra \cite{Winfinity,Winfinity2,Klebanov1}%

\begin{equation}
\int\frac{dz}{2\pi i}\psi_{J_{1},M_{1}}^{+}\psi_{J_{2},M_{2}}^{+}=(J_{2}%
M_{1}-J_{1}M_{2})\psi_{J_{1}+J_{2}-1,M_{1}+M_{2}}^{+}. \label{2D1}%
\end{equation}
This is in analogy to the work of Ref \cite{Ring,Ring1} where the ground ring
structure of ghost number zero operators was identified in the BRST
quantization. In Eq.(\ref{3.5.}), $S_{k}$ $=S_{k}\left(  \{\frac{-i\sqrt{2}%
}{k!}\partial^{k}X(0)\}\right)  $ is the Schur polynomial, which is a function
of $\{a_{k}\}=\{a_{i}:i\in\mathbb{Z}_{k}\}$ where $X$ is the conformal matter
and $\phi$ is the scalar Liouville field. We will denote the rank $(J-M)$
determinant as $\Delta(J,M,-i\sqrt{2}X)$ below.

It was remarkable that a set of discrete ZNS $G_{J,M}^{+}$ with Polyakov
momenta can be constructed \cite{ChungLee1}
\begin{equation}%
\begin{split}
G_{J,M}^{+}=  &  (J+M+1)^{-1}\int\frac{dz}{2\pi i}\left[  \psi_{1,-1}%
^{+}(z)\psi_{J,M+1}^{+}(0)+\psi_{J,M+1}^{+}(z)\psi_{1,-1}^{+}(0)\right] \\
\sim &  (J-M)!\Delta(J,M,-i\sqrt{2}X)Exp\left[  \sqrt{2}(iMX+(J-1)\phi)\right]
\\
&  +(-1)^{2J}\sum\limits_{j=1}^{J-M}(J-M-1)!\int\frac{dz}{2\pi i}%
\mathcal{D}(J,M,-i\sqrt{2}X(z),j)\\
\cdot &  Exp\left[  \sqrt{2}(i(M+1)X(z)+(J-1)\phi(z)-X(0))\right]
\end{split}
\label{3.11.}%
\end{equation}
and moreover $G_{J,M}^{+}$ were also shown \cite{ChungLee1,ChungLee2} to carry
the spacetime $\omega_{\infty}$ symmetry \cite{Winfinity,Winfinity2,Klebanov1}
charges of $2D$ string theory \cite{ChungLee1,ChungLee2}%
\begin{equation}
\int\frac{dz}{2\pi i}G_{J_{1},M_{1}}^{+}(z)G_{J_{2},M_{2}}^{+}(0)=(J_{2}%
M_{1}-J_{1}M_{2})G_{J_{1}+J_{2}-1,M_{1}+M_{2}}^{+}(0). \label{2D2}%
\end{equation}
In Eq.(\ref{3.11.}) above, $\mathcal{D}(J,M,-i\sqrt{2}X(z),j)$ is defined to
be the same as $\Delta(J,M,-i\sqrt{2}X(z))$ except that the $j^{th}$ row is
replaced by $\{(-z)^{j-1-2J},(-z)^{j-2J}...\}$. The calculation above can be
generalized to $2D$ superstring theory presented in \cite{ChungLee2}.

Moreover, it was shown that \cite{CHLTY1} the high energy limit of the
discrete ZNS $G_{J,M}^{+}$ in $2D$ string theory constructed in
Eq.(\ref{3.11.}) approaches $\psi_{J,M}^{+}$ in Eq.(\ref{3.5.}) and, as a
result, they form a high energy $w_{\infty}$ symmetry of $2D$ string theory.
This result seems to strongly suggests that the linear relations obtained from
decoupling of ZNS in $26D$ string theory are indeed closely related to the
hidden symmetry for the $26D$ string theory.

One can also use ZNS to calculate spacetime symmetries of string on compact
backgrounds. The existence of soliton ZNS at some special moduli points was
shown to be responsible for the enhanced Kac-Moody symmetry of closed bosonic
string theory. As a simple example, for the case of $26D$ bosonic closed
string compactified on a $2$-dimensional torus $T^{2}\equiv\frac{R^{2}}%
{2\pi\Lambda^{2}}$, it was pointed out that massless ZNS (including soliton
ZNS) form a representation of enhanced Kac-Moody $SU(3)_{R}\otimes$
$SU(3)_{L}$ symmetry at the moduli point \cite{Lee1}%
\begin{equation}
R_{1}=R_{2}=\sqrt{2},B=\frac{1}{2},\overset{\rightarrow}{e}_{1}=\left(
\sqrt{2},0\right)  ,\overset{\rightarrow}{e}_{2}=\left(  -\sqrt{\frac{1}{2}%
},\sqrt{\frac{3}{2}}\right)
\end{equation}
where $\Lambda^{2}$ is a $2$-dimensional lattice with a basis $\left\{
R_{1}\frac{\overset{\rightarrow}{e}_{1}}{\sqrt{2}},R_{2}\frac
{\overset{\rightarrow}{e}_{2}}{\sqrt{2}}\right\}  $, and $B$ is the
antisymmetric tensor $B_{ij}=B\epsilon_{ij}$. In the above calculation one
obtained four moduli parameters $R_{1},R_{2},B$ and $\overset{\rightarrow
}{e}_{1}\cdot\overset{\rightarrow}{e}_{2}$with $\left\vert
\overset{\rightarrow}{e}_{i}\right\vert ^{2}=2$.\ Moreover, an infinite number
of massive soliton ZNS at arbitrary higher massive level of the spectrum were
constructed in \cite{Lee1}. Presumably, these massive soliton ZNS were
responsible for enhanced stringy symmetries of the bosonic string theory.

For the case of open string compactification, unlike the closed string case
discussed above, it was pointed out that \cite{Lee2} the soliton ZNS only
exist at massive levels. These Chan-Paton soliton ZNS correspond to the
existence of the enhanced massive stringy symmetries with transformation
parameters containing both Einstein and Yang-Mills indices for the case of
Heterotic string \cite{Lee4}. On the other hand, in the T-dual picture, these
symmetries exist only at some discrete values of compactified radii when $N$
$D$-branes are coincident \cite{Lee2}.%

\setcounter{equation}{0}
\renewcommand{\theequation}{\arabic{section}.\arabic{equation}}%

\section{Stringy symmetries of hard string scattering amplitudes}

In this chapter, we will review high energy, fixed angle calculations of HSS
amplitudes. The high energy, fixed angle Ward identities derived from the
decoupling of ZNS in the HSS limit, which combines the previous two key ideas
of probing stringy symmetry, can be used to explicitly prove Gross's two
conjectures \cite{ChanLee,ChanLee1,ChanLee2, CHL, CHLTY1,CHLTY2}.

An infinite number of linear relations among high energy stringy scattering
amplitudes of different string states can be derived. Remarkably, the
algebraical constraints from these linear relations were just good enough (no
more and no less) to determine the ratios among HSS amplitudes of different
string states at each fixed mass level.

The first simple example that has been shown was the ratios among HSS
amplitudes at mass level $M^{2}=4$ \cite{ChanLee,ChanLee2} (see the definition
of polarizations $e^{T}$ and $e^{L}$ after Eq.(\ref{214}) below)
\begin{equation}
\mathcal{T}_{TTT}:\mathcal{T}_{LLT}:\mathcal{T}_{(LT)}:\mathcal{T}%
_{[LT]}=8:1:-1:-1 \label{03}%
\end{equation}
which corresponds to stringy symmetries in the $\sigma$-model calculation
discussed from Eq.(\ref{BZNS}) to Eq.(\ref{02}). Eq.(\ref{03}) is valid at any
order in string perturbation theory since we expect the decoupling of ZNS to
be valid for arbitrary string loop amplitudes \cite{ChanLee3}.

One of the three methods to calculate Eq.(\ref{03}) is to use the method of
decoupling of ZNS. We first note that there are four ZNS at mass level
\ $M^{2}$ $=4$ in the old covariant first quantized string spectrum. For type
I ZNS, there is one symmetric spin two tensor, one vector and one scalar ZNS.
In addition, it is important to note that there exists one vector type II ZNS.
The corresponding Ward identities for these four ZNS were calculated to be
\cite{JCLee}%

\begin{align}
k_{\mu}\theta_{\nu\lambda}\mathcal{T}_{\chi}^{(\mu\nu\lambda)}+2\theta_{\mu
\nu}\mathcal{T}_{\chi}^{(\mu\nu)}  &  =0,\\
\left(  \frac{5}{2}k_{\mu}k_{\nu}\theta_{\lambda}^{\prime}+\eta_{\mu\nu}%
\theta_{\lambda}^{\prime}\right)  \mathcal{T}_{\chi}^{(\mu\nu\lambda)}%
+9k_{\mu}\theta_{\nu}^{\prime}\mathcal{T}_{\chi}^{(\mu\nu)}+6\theta_{\mu
}^{\prime}\mathcal{T}_{\chi}^{\mu}  &  =0,\\
\left(  \frac{1}{2}k_{\mu}k_{\nu}\theta_{\lambda}+2\eta_{\mu\nu}%
\theta_{\lambda}\right)  \mathcal{T}_{\chi}^{(\mu\nu\lambda)}+9k_{\mu}%
\theta_{\nu}\mathcal{T}_{\chi}^{[\mu\nu]}-6\theta_{\mu}\mathcal{T}_{\chi}%
^{\mu}  &  =0,\label{D22}\\
\left(  \frac{17}{4}k_{\mu}k_{\nu}k_{\lambda}+\frac{9}{2}\eta_{\mu\nu
}k_{\lambda}\right)  \mathcal{T}_{\chi}^{(\mu\nu\lambda)}+(9\eta_{\mu\nu
}+21k_{\mu}k_{\nu})\mathcal{T}_{\chi}^{(\mu\nu)}+25k_{\mu}\mathcal{T}_{\chi
}^{\mu}  &  =0,
\end{align}
where the polarization $\theta_{\mu\nu}$ is a transverse and traceless tensor,
$\theta_{\lambda}^{\prime}$ and $\theta_{\lambda}$ are transverse vectors.
$\mathcal{T}_{\chi}^{\prime}s$ in the above equations are the mass level
$M^{2}$ $=4$, $\chi$-th order string-loop amplitudes. In the above equations,
$v_{2}(k_{2})$\ is chosen to be the physical vertex operators constructed from
ZNS and $k_{\mu}\equiv k_{2\mu}$. Note that Eq.(\ref{D22}) is the
"inter-particle Ward identity" corresponding to $D_{2}$ vector ZNS in
Eq.(\ref{02}) obtained by antisymmetrizing the terms containing $\alpha
_{-1}^{\mu}\alpha_{-2}^{\nu}$ \cite{Lee}. We will use $1$ and $2$ to represent
the incoming particles and $3$ and $4$ for the scattered particles. In the
Ward identities calculated above, the vertices can be any string states and
for simplicity we have omitted their tensor indices for the cases of excited
string states for\ the vertices $1$, $3$ and $4$.

In the HSS limit, one enjoys many simplifications in the calculation. First of
all, all polarizations of the scattering amplitudes that are orthogonal to the
scattering plane are of subleading order in energy, and one needs only
consider polarizations on the scattering plane. Secondly, to the leading order
in energy, $e^{P}\simeq$ $e^{L}$ in the HSS calculation. In the end of the
calculation, one ends up with the simple linear equations for leading order
HSS amplitudes \cite{ChanLee,ChanLee2}%

\begin{align}
\mathcal{T}_{LLT}^{5\rightarrow3}+\mathcal{T}_{(LT)}^{3}  &  =0,\label{212}\\
10\mathcal{T}_{LLT}^{5\rightarrow3}+\mathcal{T}_{TTT}^{3}+18\mathcal{T}%
_{(LT)}^{3}  &  =0,\label{213}\\
\mathcal{T}_{LLT}^{5\rightarrow3}+\mathcal{T}_{TTT}^{3}+9\mathcal{T}%
_{[LT]}^{3}  &  =0 \label{214}%
\end{align}
where $e^{P}=\frac{1}{M_{2}}(E_{2},\mathrm{k}_{2},0)=\frac{k_{2}}{M_{2}}$ the
momentum polarization, $e^{L}=\frac{1}{M_{2}}(\mathrm{k}_{2},E_{2},0)$ the
longitudinal polarization and $e^{T}=(0,0,1)$ the transverse polarization are
three polarizations on the scattering plane. In Eq.(\ref{212}) to
Eq.(\ref{214}), each scattering amplitude has been assigned a relative energy
power. For each longitudinal $L$ component, the energy order is $E^{2}$ while
for each transverse $T$ component, the energy order is $E.$ This is due to the
definitions of $e_{L}$and $e_{T}$ above, where $e_{L}$ got one energy power
more than that of $e_{T}.$ By Eq.(\ref{213}), the naive leading order $E^{5}$
term of the energy expansion for $\mathcal{T}_{LLT}$ is forced to be zero. As
a result, the real leading order term is $E^{3}$. Similar rule also applies to
$\mathcal{T}_{LLT}$ in Eq.(\ref{212}) and Eq.(\ref{214}). The solution of
these three linear relations gives Eq.(\ref{03}). Eq.(\ref{03}) gives the
first evidence of Gross conjecture \cite{Gross,Gross1} on HSS amplitudes.

To confirm the validity of the calculation of decoupling of ZNS above, a
sample calculation of HSS amplitudes for mass level $M^{2}$ $=4$
\cite{ChanLee2} justified the ratios calculated in Eq.(\ref{03}). Since the
ratios in Eq.(\ref{03}) are independent of the choices of string states at the
vertices $1$, $3$ and $4$, for simplicity, we will choose them to be all
tachyons. At the string-tree level $\chi=1$, with a tensor string state at
vertex $2$ and three tachyons at the vertices $1$, $3$ and $4$, all HSS
amplitudes of mass level $M_{2}^{2}$ $=4$ can be calculated to be ($s-t$
channel)%
\begin{equation}
\mathcal{T}_{TTT}=-8E^{9}\mathcal{T}(3)\sin^{3}\phi_{CM}\left[  1+\frac
{3}{E^{2}}+\frac{5}{4E^{4}}-\frac{5}{4E^{6}}+O\left(  \frac{1}{E^{8}}\right)
\right]  ,
\end{equation}

\begin{align}
\mathcal{T}_{LLT}  &  =-E^{9}\mathcal{T}(3)\left[  \sin^{3}\phi_{CM}%
+(6\sin\phi_{CM}\cos^{2}\phi_{CM})\frac{1}{E^{2}}\right. \nonumber\\
&  \left.  -\sin\phi_{CM}\left(  \frac{11}{2}\sin^{2}\phi_{CM}-6\right)
\frac{1}{E^{4}}+O\left(  \frac{1}{E^{6}}\right)  \right]  ,
\end{align}

\begin{align}
\mathcal{T}_{[LT]}  &  =E^{9}\mathcal{T}(3)\left[  \sin^{3}\phi_{CM}%
-(2\sin\phi_{CM}\cos^{2}\phi_{CM})\frac{1}{E^{2}}\right. \nonumber\\
&  \left.  +\sin\phi_{CM}\left(  \frac{3}{2}\sin^{2}\phi_{CM}-2\right)
\frac{1}{E^{4}}+O\left(  \frac{1}{E^{6}}\right)  \right]  ,
\end{align}

\begin{align}
\mathcal{T}_{(LT)}  &  =E^{9}\mathcal{T}(3)\left[  \sin^{3}\phi_{CM}+\sin
\phi_{CM}\left(  \frac{3}{2}-10\cos\phi_{CM}-\frac{3}{2}\cos^{2}\phi
_{CM}\right)  \frac{1}{E^{2}}\right. \nonumber\\
&  \left.  -\sin\phi_{CM}\left(  \frac{1}{4}+10\cos\phi_{CM}+\frac{3}{4}%
\cos^{2}\phi_{CM}\right)  \frac{1}{E^{4}}+O\left(  \frac{1}{E^{6}}\right)
\right]  ,
\end{align}
where%
\begin{align}
\mathcal{T}(N)  &  \mathcal{=}\sqrt{\pi}(-1)^{N-1}2^{-n}E^{-1-2N}\left(
\sin\frac{\phi_{CM}}{2}\right)  ^{-3}\left(  \cos\frac{\phi_{CM}}{2}\right)
^{5-2N}\nonumber\\
&  \cdot\exp\left(  -\frac{s\ln s+t\ln t-(s+t)\ln(s+t)}{2}\right)  ,
\end{align}
is the high energy limit of $\frac{\Gamma(-\frac{s}{2}-1)\Gamma(-\frac{t}%
{2}-1)}{\Gamma(\frac{u}{2}+2)}$ with $s+t+u=2N-8$. We thus have justified the
ratios calculated in Eq.(\ref{03}) with $\mathcal{T}_{TTT}^{3}=-8E^{9}%
\mathcal{T}(3)\sin^{3}\phi_{CM}$.

The calculations based on ZNS thus relate \cite{CLYang} gauge transformation
of WSFT to high energy string symmetries of Gross. However, in the sample
calculation of \cite{GrossManes}, two of the four high energy amplitudes in
Eq.(\ref{03}) were missing, and as a result the decoupling of ZNS or unitarity
was violated there. This is due to the unawareness of the importance of ZNS in
the saddle-point calculation of \cite{GM,GM1, Gross, Gross1,GrossManes}.

The calculations for $M^{2}$ $=4$ above can be generalized to $M^{2}$ $=6$
\cite{ChanLee2}. To the leading order in the hard scattering limit, one ended
up with $8$ constraint equations and $9$ HSS amplitudes. A calculation showed
that \cite{ChanLee2}%

\begin{align}
\mathcal{T}_{TTTT}^{4}  &  :\mathcal{T}_{TTLL}^{4}:\mathcal{T}_{LLLL}%
^{4}:\mathcal{T}_{TTL}^{4}:\mathcal{T}_{LLL}^{4}:\widetilde{\mathcal{T}%
}_{LT,T}^{4}:\widetilde{\mathcal{T}}_{LP,P}^{4}:\mathcal{T}_{LL}%
^{4}:\widetilde{\mathcal{T}}_{LL}^{4}=\nonumber\\
16  &  :\frac{4}{3}:\frac{1}{3}:-\frac{4\sqrt{6}}{9}:-\frac{\sqrt{6}}%
{9}:-\frac{2\sqrt{6}}{3}:0:\frac{2}{3}:0. \label{055}%
\end{align}
The results for mass level $M^{2}$ $=8$ \cite{CHLTY1} can also be obtained
with more lengthy calculation.

Remarkably, the above results for up to mass level $M^{2}$ $=8$ can be
generalized to arbitrary higher mass levels. From the calculations of
Eq.(\ref{212}) to Eq.(\ref{214}), one first observes that only string states
of the following form \cite{CHLTY1,CHLTY2}
\begin{equation}
\left\vert N,2m,q\right\rangle \equiv(\alpha_{-1}^{T})^{N-2m-2q}(\alpha
_{-1}^{L})^{2m}(\alpha_{-2}^{L})^{q}|0,k\rangle\label{Nmq}%
\end{equation}
are of leading order in energy in the HSS limit. The request of only even
power $2m$ in $\alpha_{-1}^{L}$ is the result from the observation that the
naive energy order of the HSS amplitudes will drop by even number of energy
power as can be seen in Eq.(\ref{212}) to Eq.(\ref{214}). It can be shown that
the HSS amplitudes for states with $(\alpha_{-1}^{L})^{2m+1}$\ are of
subleading order in energy and can be ignored at the beginning of the
calculation. The Ward identities could be simplified a lot if we only consider
the high energy states in Eq.(\ref{Nmq}) in the HSS limit. First, consider the
decoupling of type I \ high energy ZNS
\begin{equation}
L_{-1}|N-1,2m-1,q\rangle\simeq M|N,2m,q\rangle+(2m-1)|N,2m-2,q+1\rangle
\end{equation}
where the terms that are not in the form of Eq.(\ref{Nmq}) can be omitted.
This implies that
\begin{equation}
\mathcal{T}^{(N,2m,q)}=-\frac{2m-1}{M}\mathcal{T}^{(N,2m-2,q+1)}.
\end{equation}
Using this recurrent relation, one can easily obtain
\begin{equation}
\mathcal{T}^{(N,2m,q)}=\frac{(2m-1)!!}{(-M)^{m}}\mathcal{T}^{(N,0,m+q)}
\label{123}%
\end{equation}
where the double factorial is defined to be $(2m-1)!!=\frac{(2m)!}{2^{m}m!}$.

Next, we consider the decoupling of type II high energy ZNS
\begin{equation}
L_{-2}|N-2,0,q\rangle\simeq\frac{1}{2}|N,0,q\rangle+M|N,0,q+1\rangle.
\end{equation}
Similarly, the irrelevant terms of the subleading order in energy can be
omitted, and it implies that%
\begin{equation}
\mathcal{T}^{(N,0,q+1)}=-\frac{1}{2M}\mathcal{T}^{(N,0,q)},
\end{equation}
which leads to the relation
\begin{equation}
\mathcal{T}^{(N,0,q)}=\frac{1}{(-2M)^{q}}\mathcal{T}^{(N,0,0)}. \label{124}%
\end{equation}

Finally, the ratios for arbitrary mass levels $M^{2}=2(N-1)$ is an immediate
deduction of the above two equations, Eq.(\ref{123}) and Eq.(\ref{124}),
\cite{CHLTY1,CHLTY2}
\begin{equation}
\frac{T^{(N,2m,q)}}{T^{(N,0,0)}}=\left(  -\frac{1}{M}\right)  ^{2m+q}\left(
\frac{1}{2}\right)  ^{m+q}(2m-1)!!. \label{04}%
\end{equation}

To justify the ratios calculated in Eq.(\ref{04}) by the method of decoupling
of high energy ZNS, it was shown that exactly the same results can also be
consistently obtained by two other calculations, the Virasoro constraint
calculation and the saddle-point calculation. Here we review the saddle-point
calculation. Since the result in Eq.(\ref{04}) is valid for all string loop
order, we need only do saddle-point calculation of the string tree level
scattering amplitudes. Without loss of generality, we will choose the vertices
$1$,$3$ and $4$ to be tachyons, and the vertex $2$ to be in the form of
Eq.(\ref{Nmq}). The $t-u$ channel contribution to the stringy amplitude at
tree level is
\begin{align}
\mathcal{T}^{(N,2m,q)}  &  =\int_{1}^{\infty}dxx^{(1,2)}(1-x)^{(2,3)}\left[
\frac{e^{T}\cdot k_{1}}{x}-\frac{e^{T}\cdot k_{3}}{1-x}\right]  ^{N-2m-2q}%
\nonumber\\
&  \cdot\left[  \frac{e^{P}\cdot k_{1}}{x}-\frac{e^{P}\cdot k_{3}}%
{1-x}\right]  ^{2m}\left[  -\frac{e^{P}\cdot k_{1}}{x^{2}}-\frac{e^{P}\cdot
k_{3}}{(1-x)^{2}}\right]  ^{q}%
\end{align}
where we have defined the notation $(1,2)=k_{1}\cdot k_{2}$ etc.

In the saddle-point calculation, we transform the above scattering amplitude
into the following form \cite{CHLTY1,CHLTY2}%
\begin{equation}
\mathcal{T}^{(N,2m,q)}(K)=\int_{1}^{\infty}dx\mbox{ }u(x)e^{-Kf(x)}%
\end{equation}
where various quantities above are defined to be
\begin{align}
K  &  \equiv-(1,2)\rightarrow2E^{2},\\
\tau &  \equiv-\frac{(2,3)}{(1,2)}\rightarrow\sin^{2}\frac{\phi}{2},\\
f(x)  &  \equiv\ln x-\tau\ln(1-x),\\
u(x)  &  \equiv\left[  \frac{(1,2)}{M}\right]  ^{2m+q}(1-x)^{-N+2m+2q}%
(f^{\prime})^{2m}(f^{\prime\prime})^{q}(-e^{T}\cdot k_{3})^{N-2m-2q}.
\end{align}
The saddle-point for the integration, $x=x_{0}$, is defined to be
\begin{equation}
f^{\prime}(x_{0})=0,
\end{equation}
and we have%
\begin{equation}
x_{0}=\frac{1}{1-\tau}\text{ \ ,}\hspace{1cm}f^{\prime\prime}(x_{0}%
)=(1-\tau)^{3}\tau^{-1}. \label{saddle}%
\end{equation}
It is very crucial to note that%
\begin{equation}
u(x_{0})=u^{\prime}(x_{0})=....=u^{(2m-1)}(x_{0})=0,
\end{equation}
and the leading term can be calculated to be
\begin{equation}
u^{(2m)}(x_{0})=\left[  \frac{(1,2)}{M}\right]  ^{2m+q}(1-x_{0})^{-N+2m+2q}%
(2m)!(f_{0}^{\prime\prime})^{2m+q}(-e^{T}\cdot k_{3})^{N-2m-2q}.
\end{equation}

One can now calculate the Gaussian integral associated with the four-point HSS
amplitudes
\begin{align}
&  \int_{1}^{\infty}dx\mbox{ }u(x)e^{-Kf(x)}\nonumber\\
&  =\sqrt{\frac{2\pi}{Kf_{0}^{\prime\prime}}}e^{-Kf_{0}}\left[  \frac
{u_{0}^{(2m)}}{2^{m}\ m!\ (f_{0}^{\prime\prime})^{m}\ K^{m}}+O(\frac
{1}{K^{m+1}})\right] \nonumber\\
&  =\sqrt{\frac{2\pi}{Kf_{0}^{\prime\prime}}}e^{-Kf_{0}}\left[  (-1)^{N-q}%
\frac{2^{N-2m-q}(2m)!}{m!\ {M}^{2m+q}}\ \tau^{-\frac{N}{2}}(1-\tau)^{\frac
{3N}{2}}E^{N}+O(E^{N-2})\right]  .
\end{align}
This result explicitly shows that with one higher spin tensor and three
tachyons, the four-point HSS amplitudes have the same dependence on the
scattering angle at each mass level $N$. One can then extract the ratios
\begin{align}
\lim_{E\rightarrow\infty}\frac{\mathcal{T}^{(N,2m,q)}}{\mathcal{T}^{(N,0,0)}}
&  =\frac{(-1)^{q}(2m)!}{m!(2M)^{2m+q}}\nonumber\\
&  =\left(  -\frac{2m-1}{M}\right)  ....\left(  -\frac{3}{M}\right)  \left(
-\frac{1}{M}\right)  \left(  -\frac{1}{2M}\right)  ^{m+q},
\end{align}
which is consistent with the calculation of decoupling of high energy ZNS
obtained in Eq.(\ref{04}).

We conclude that at each fixed mass level there is only one independent HSS
amplitude. As a result, one can then express the general four-point HSS
amplitude for four arbitrary string states in terms of four tachyon scattering
amplitude. This completes the proof \cite{ChanLee,ChanLee1,ChanLee2, CHL,
CHLTY1,CHLTY2} of Gross's two conjectures on high energy symmetry of string
theory stated above.

All the above calculations can be generalized to the case of the NS-sector of
hard superstring scattering amplitudes. However, it was interesting to find
that \cite{susy} there were new HSS amplitudes for the superstring case. The
worldsheet fermion exchange in the correlation functions induces new
contributions to the high energy scattering amplitudes of string states with
polarizations orthogonal to the scattering plane. This is presumably related
to the high energy massive spacetime fermionic scattering amplitudes in the
R-sector of superstring theory.

Incidentally, it was important to discover \cite{ChanLee,ChanLee1,ChanLee2,
CHL} that the result of saddle-point calculation in Refs \cite{GM,GM1,
Gross,Gross1, GrossManes} was inconsistent with the high energy ZNS
calculation in Refs \cite{ChanLee,ChanLee1,ChanLee2, CHL}. A simple example is
that two of the four amplitudes in Eq.(\ref{03})\ were missed as was pointed
out in \cite{ChanLee,ChanLee2}. The \textit{corrected saddle-point
calculation} was given in \cite{CHL} where the missing terms in Refs
\cite{GM,GM1, Gross,Gross1,GrossManes} were remedied to recover the stringy
Ward identities.

Indeed, it was found \cite{CHL} that saddle point calculation in \cite{GM,GM1,
Gross,Gross1,GrossManes} is only valid for the four tachyon amplitude. In
general, the results calculated in \cite{GM,GM1, Gross,Gross1,GrossManes} gave
the right energy exponent in the scattering amplitudes, but not the energy
power factors in front of the exponential for the cases of the massive
\textit{excited string states}. These energy power factors are subleading
terms ignored in \cite{GM,GM1, Gross,Gross1,GrossManes} but they are crucial
if one wants to get the linear relations among HSS amplitudes conjectured by Gross.

Interestingly, the inconsistency of the saddle point calculation discussed
above for the excited string states was also pointed out by the authors of
\cite{West2}. The source of disagreement in their group theoretic approach of
stringy symmetries stems from the proper choice of local coordinates for the
worldsheet saddle points to describe the behavior of the excited string states
at high energy limit. It seems that both the ZNS calculation and the
calculation based on group theoretic approach agree with tachyon amplitudes
obtained in \cite{GM,GM1, Gross,Gross1, GrossManes} (ignore the possible phase
factors in the amplitudes to be discussed in the next chapter), but disagree
with amplitudes for other excited string states.%

\setcounter{equation}{0}
\renewcommand{\theequation}{\arabic{section}.\arabic{equation}}%

\section{Hard closed string scatterings, KLT and hard string BCJ relations}

The next interesting issues are the calculation of \textit{closed} string
scattering amplitudes and their symmetries in the HSS limit \cite{Closed}.
Historically, the open string four tachyon amplitude in the HSS limit was
first calculated in the original paper of Veneziano in 1968. On the other
hand, the $\mathcal{N}$-loop closed HSS amplitudes were calculated by the
saddle-point method in 1988 \cite{GM,GM1}. Both open and closed HSS amplitudes
exhibit the very soft exponential fall-off behaviors in contrast to the power
law behavior of the scattering amplitudes of quantum field theory.

However, an inconsistence would arise if one simply plugs, for example, the
tree level four tachyon open and closed string HSS amplitudes\ calculated by
these authors into the KLT relation \cite{KLT}%
\begin{equation}
A_{\text{closed}}^{\left(  4\right)  }\left(  s,t,u\right)  =\sin\left(  \pi
k_{2}\cdot k_{3}\right)  A_{\text{open}}^{\left(  4\right)  }\left(
s,t\right)  \bar{A}_{\text{open}}^{\left(  4\right)  }\left(  t,u\right)
\label{KLT2}%
\end{equation}
which is valid for \textit{all} kinematic regimes and for \textit{all} string
states. This inconsistence is due to the phase factor $\sin\left(  \pi
k_{2}\cdot k_{3}\right)  $ in the above equation which was missing in the
closed string saddle-point calculation in \cite{GM,GM1}. One simple clue to
see the origin of this inconsistence is to note that the saddle-point
$x_{0}=\frac{1}{1-\tau}>1$ identified for the open string calculation in
Eq.(\ref{saddle}) is in the regime $[1,\infty)$. So only saddle point
calculation for $\bar{A}_{\text{open}}^{\left(  4\right)  }\left(  t,u\right)
$ is reliable, but not that of $A_{\text{open}}^{\left(  4\right)  }\left(
s,t\right)  $ and neither that of closed string amplitude $A_{\text{closed}%
}^{\left(  4\right)  }\left(  s,t,u\right)  $ \cite{Closed} by the KLT relation.

Instead of using saddle-point calculation for the closed HSS amplitudes, the
above considerations led the authors of \cite{Closed} to study the
relationship between $A_{\text{open}}^{\left(  4\right)  }\left(  s,t\right)
$ and $\bar{A}_{\text{open}}^{\left(  4\right)  }\left(  t,u\right)  $ for
arbitrary string states in the HSS limit. With the help of the
\textit{infinite linear relations in Eq.(\ref{04})}, one needs only calculate
relationship between $s-t$ and $t-u$ channel HSS amplitudes for the leading
trajectory string states which were much easier to calculate. They ended up
with the following result in the HSS limit (2006) \cite{Closed}%
\begin{equation}
A_{\text{open}}^{\left(  4\right)  }\left(  s,t\right)  =\frac{\sin\left(  \pi
k_{2}.k_{4}\right)  }{\sin\left(  \pi k_{1}k_{2}\right)  }\bar{A}%
_{\text{open}}^{\left(  4\right)  }\left(  t,u\right)  \text{ \ }(HSS),
\label{CLY-BCJ}%
\end{equation}
which is valid for four arbitrary string states. It is now clear that due to
the phase factor in the above equation, the saddle-point calculation of
$A_{\text{open}}^{\left(  4\right)  }\left(  s,t\right)  $ is not reliable,
neither for the closed one $A_{\text{closed}}^{\left(  4\right)  }\left(
s,t,u\right)  $ in view of the KLT relation in Eq.(\ref{KLT2}). One can now
use the reliable saddle-point calculation of $\bar{A}_{\text{open}}^{\left(
4\right)  }\left(  t,u\right)  $%
\begin{equation}
A_{\text{open}}^{(4-\text{tachyon})}\left(  t,u\right)  \simeq(stu)^{-\frac
{3}{2}}\exp\left(  -\frac{s\ln s+t\ln t+u\ln u}{2}\right)  , \label{exp}%
\end{equation}
and Eq.(\ref{CLY-BCJ}) to calculate $A_{\text{open}}^{\left(  4\right)
}\left(  s,t\right)  $ in the HSS limit. The consistent closed string
four-tachyon HSS amplitudes can then be calculated by using the KLT relation
in Eq.(\ref{KLT2}) to be \cite{Closed}%
\begin{equation}
A_{\text{closed}}^{(4-\text{tachyon})}\left(  s,t,u\right)  \simeq\frac
{\sin\left(  \pi t/2\right)  \sin\left(  \pi u/2\right)  }{\sin\left(  \pi
s/2\right)  }(stu)^{-3}\exp\left(  -\frac{s\ln s+t\ln t+u\ln u}{4}\right)  .
\label{open1}%
\end{equation}

The exponential factor in Eq.(\ref{exp}) was first discussed by Veneziano
\cite{Veneziano}. The result for the high energy closed string four-tachyon
amplitude in Eq.(\ref{open1}) differs from the one calculated in the
literature \cite{GM,GM1} by an oscillating factor $\frac{\sin\left(  \pi
t/2\right)  \sin\left(  \pi u/2\right)  }{\sin\left(  \pi s/2\right)  }$. It
is important to note that the results of Eqs.(\ref{open1}), (\ref{exp}) and
Eq.(\ref{CLY-BCJ}) are consistent with the KLT formula, while the calculation
in \cite{GM,GM1} is NOT.

Indeed, one might try to use the saddle-point method to calculate the high
energy closed string scattering amplitude. The closed string four-tachyon
scattering amplitude is%
\begin{align}
A_{\text{closed}}^{(4-\text{tachyon})}\left(  s,t,u\right)   &  =\int
dxdy\exp\left(  \frac{k_{1}\cdot k_{2}}{2}\ln\left\vert z\right\vert
+\frac{k_{2}\cdot k_{3}}{2}\ln\left\vert 1-z\right\vert \right) \nonumber\\
&  \equiv\int dxdy(x^{2}+y^{2})^{-2}[\left(  1-x\right)  ^{2}+y^{2}]^{-2}%
\exp\left[  -Kf\left(  x,y\right)  \right]
\end{align}
where $K=s/8$ and $f(x,y)=\ln(x^{2}+y^{2})-\tau\ln[(1-x)^{2}+y^{2}]$ with
$\tau=-t/s$. The "saddle-point" of $\ f(x,y)$ can then be calculated to be%
\begin{equation}
\nabla f\left(  x,y\right)  \mid_{x_{0}=\frac{1}{1-\tau},y_{0}=0}=0.
\end{equation}
The HSS limit of the closed string four-tachyon scattering amplitude is
calculated to be%
\begin{align}
A_{\text{closed}}^{(4-\text{tachyon})}\left(  s,t,u\right)   &  \simeq
\frac{2\pi}{K\sqrt{\det\frac{\partial^{2}f(x_{0},y_{0})}{\partial x\partial
y}}}\exp[-Kf(x_{0},y_{0})]\nonumber\\
&  \simeq(stu)^{-3}\exp\left(  -\frac{s\ln s+t\ln t+u\ln u}{4}\right)  ,
\end{align}
which is consistent with the result calculated in the literature
\cite{GM,GM1}, but is different from the one in Eq.(\ref{open1}). However, one
notes that%
\begin{equation}
\frac{\partial^{2}f(x_{0},y_{0})}{\partial x^{2}}=\frac{2(1-\tau)^{3}}{\tau
}=-\frac{\partial^{2}f(x_{0},y_{0})}{\partial y^{2}},\frac{\partial^{2}%
f(x_{0},y_{0})}{\partial x\partial y}=0,
\end{equation}
which means that $(x_{0},y_{0})$ is NOT the local minimum of $f(x,y)$, and one
should not trust this saddle-point calculation. There was other evidence
pointed out by authors of \cite{Closed} to support this conclusion. Finally,
the ratios of closed HSS amplitudes turned out to be the tensor products of
two open string ratios
\begin{equation}
\frac{T^{\left(  N;2m,2m^{^{\prime}};q,q^{^{\prime}}\right)  }}{T^{\left(
N;0,0;0,0\right)  }}=\left(  -\frac{1}{M_{2}}\right)  ^{2\left(
m+m^{^{\prime}}\right)  +q+q^{^{\prime}}}\left(  \frac{1}{2}\right)
^{m+m^{^{\prime}}+q+q^{^{\prime}}}(2m-1)!!(2m^{\prime}-1)!!. \label{closed4}%
\end{equation}

The relationship between $s-t$ and $t-u$ channels HSS amplitudes in
Eq.(\ref{CLY-BCJ}) was later argued to be valid for \textit{all} kinematic
regime (2009) \cite{BCJ2}%
\begin{equation}
A_{\text{open}}^{\left(  4\right)  }\left(  s,t\right)  =\frac{\sin\left(  \pi
k_{2}.k_{4}\right)  }{\sin\left(  \pi k_{1}k_{2}\right)  }\bar{A}%
_{\text{open}}^{\left(  4\right)  }\left(  t,u\right)  , \label{SBCJ}%
\end{equation}
the so-called "string BCJ relation", based on the monodromy of integration in
the string amplitudes calculation in 2009 \cite{BCJ2}. An explicit proof of
Eq.(\ref{CLY-BCJ}) for arbitrary four string states and all kinematic regimes
was given recently in \cite{LLY1,LLY2}.

The motivation for the author in \cite{BCJ2} to calculate Eq.(\ref{SBCJ}) was
different from the result in Eq.(\ref{CLY-BCJ}) which was motivated by the
calculation of hard closed string scattering amplitudes. While the motivation
in \cite{BCJ2} was based on the field theory BCJ relation \cite{BCJ1} for the
scattering amplitudes $A$ in Yang-Mills theory, which was first pointed out
and calculated in 2008 \cite{BCJ1} to be
\begin{equation}
sA(k_{1},k_{2},k_{3},k_{4})-uA(k_{1},k_{4},k_{2},k_{3})=0. \label{fbcj}%
\end{equation}
Note that for the supersymmetric case, there is no tachyon and the low energy
massless limit of Eq.(\ref{SBCJ}) reproduces Eq.(\ref{fbcj}).

Recently, the stringy generalization of the massless field theory BCJ to the
higher spin string states was calculated to be \cite{LLY1,LLY2}%
\begin{equation}
\frac{A_{st}^{(p,r,q)}}{A_{tu}^{(p,r,q)}}=\left(  -1\right)  ^{N}%
\frac{B\left(  -M_{1}M_{2}+1,\frac{M_{1}M_{2}}{2}\right)  }{B\left(
\frac{M_{1}M_{2}}{2},\frac{M_{1}M_{2}}{2}\right)  }\simeq\frac{\sin\pi\left(
k_{2}\cdot k_{4}\right)  }{\sin\pi\left(  k_{1}\cdot k_{2}\right)  }
\label{level}%
\end{equation}
by taking the \textit{nonrelativistic} limit $|\vec{k_{2}}|<<M_{S}$ of
Eq.(\ref{CLY-BCJ}). In Eq.(\ref{level}), $B$ was the beta function, and
$k_{1}$, $k_{3}$ and $k_{4}$ were taken to be tachyons, and $k_{2}$ was the
following tensor string state%

\begin{equation}
V_{2}=(i\partial X^{T})^{p}(i\partial X^{L})^{r}(i\partial X^{P})^{q}%
e^{ik_{2}X}%
\end{equation}
where%
\begin{equation}
N=p+r+q\text{, \ }M_{2}^{2}=2(N-1)\text{, }N\geq2.
\end{equation}
Moreover, many \textit{reduced recurrence relations in the}
\textit{nonrelativistic string scattering (NSS) limit} can be derived
\cite{LLY1,LLY2}. These recurrence relations relates nonrelativistic string
scattering amplitudes of different higher spin particles at both $s-t$ and
$t-u$ channels for each fixed mass level.

The generalization of the four point function relation in Eq.(\ref{CLY-BCJ})
to higher point string amplitudes can be found in \cite{BCJ2}. It is
interesting to see that historically the four point (high energy) "string BCJ
relations" Eq.(\ref{CLY-BCJ}) \cite{Closed} were discovered even earlier than
the field theory BCJ relations in Eq.(\ref{fbcj})! \cite{BCJ1} (and earlier
than string BCJ relations in Eq.(\ref{SBCJ}) \cite{BCJ2}).

The ratios calculated in Eq.(\ref{closed4}) persist for the case of closed
string D-particle scatterings in the HSS limit. For the simple case of
$m=0=m^{\prime}$, the ratios were first calculated to be $\left(  -\frac
{1}{2M}\right)  ^{q+q^{\prime}}$ \cite{Dscatt}. The complete ratios were\ then
calculated through the open string HSS ratios Eq.(\ref{04}), and were found to
be \textit{factorized} \cite{LMY}
\begin{equation}
\frac{T_{SD}^{\left(  N;2m,2m^{^{\prime}};q,q^{^{\prime}}\right)  }}%
{T_{SD}^{\left(  N;0,0;0,0\right)  }}=\left(  -\frac{1}{M_{2}}\right)
^{2\left(  m+m^{^{\prime}}\right)  +q+q^{^{\prime}}}\left(  \frac{1}%
{2}\right)  ^{m+m^{^{\prime}}+q+q^{^{\prime}}}(2m-1)!!(2m^{\prime}-1)!!.
\label{corresp2}%
\end{equation}

It has been known for a long time that the closed strings scattering
amplitudes can be factorized into two open strings scattering amplitudes due
to the existence of the KLT formula \cite{KLT}. On the contrary, there is no
physical picture for the amplitude of open strings scattered from a D-particle
and thus no factorization for the amplitude of closed strings scattered from a
D-particle into two channels of open strings scattered from a D-particle, and
hence no KLT-like formula in that case. Therefore, the factorized ratios in
HSS regime calculated above came as a surprise. However, these ratios are
consistent with the decoupling of high energy ZNS calculated previously in
\cite{ChanLee,ChanLee1,ChanLee2, CHL,CHLTY1,CHLTY2,CHLTY3,susy,Closed}. It
will be helpful to clarify this paradox if one can calculate the complete HSS
amplitudes and see how the \textit{non-factorized amplitudes} can give the
result of factorized ratios.

On the other hand, it was shown that the HSS amplitudes of closed string
scattered from D24-brane, or D-domain-wall, behave as \textit{power-law with
Regge-pole structure }\cite{Wall} instead of the exponential fall-off
behavior. This is to be compared with the well-known power law behavior of the
D-instanton scatterings.

This discovery makes D-domain-wall scatterings an unique example of a hybrid
of string and field theory scatterings. Moreover, it was discovered that
\cite{Wall} the usual linear relations of HSS amplitudes of Eq.(\ref{corresp2}%
) breaks down for the D-domain-wall scatterings. This gives a strong evidence
that the existence of the infinite linear relations, or stringy symmetries, of
HSS amplitudes is responsible for the softer, exponential fall-off behavior of
HSS scatterings than the power-law field theory scatterings.

Being a consistent theory of quantum gravity, string theory is remarkable for
its soft ultraviolet structure. This is mainly due to \textit{three} closely
related fundamental characteristics of HSS amplitudes. The first is the
exponential fall-off behavior of the HSS amplitudes in contrast to the
power-law field theory scatterings amplitudes. The second is the existence of
infinite Regge poles in the string scattering amplitudes. The existence of
infinite linear relations discussed in chapter III of this review is the
\textit{third} fundamental characteristics of HSS amplitudes.

It will be important to study more string scatterings to justify the above
three fundamental characteristics in the HSS limit. The scatterings of
massless states from Orientifold planes have been studied in the literature by
using the boundary states formalism \cite{Craps,Craps1,Craps2,Schnitzer}, and
on the worldsheet of real projected plane $RP_{2}$ \cite{Garousi}. Many
speculations have been made about the scatterings of \textit{massive} string
states from the O-domain-wall scatterings. It is one of the purposes of
reference \cite{O-plane} to clarify these speculations and to discuss their
relations with the above three fundamental characteristics of HSS scatterings.
In \cite{O-plane}, the authors studied closed strings scattered from O-planes.
In particular they calculated massive closed string states at arbitrary mass
levels scattered from Orientifold planes in the HSS limit. Except for
O-domain-wall, as we explained above, one expects the infinite linear
relations HSS for the generic O$p$-planes with $p\geq0$. For simplicity, one
considered only the case of O-particle HSS \cite{O-plane}, and confirmed that
there exist only $t$-channel closed string Regge poles in the form factor of
the O-particle scatterings amplitudes as expected.

Like the well-known D-instanton scatterings, one found that the amplitudes of
O-domain-wall scatterings behave like field theory scatterings, namely
\textit{UV power-law without Regge pole}. In addition, only finite number of
$t$-channel closed string poles in the O-domain-wall scatterings was found
with the masses of the poles being bounded by the masses of the external legs
\cite{O-plane}. One thus confirmed that all massive closed string states do
couple to the O-domain-wall as was conjectured previously \cite{Myers,Garousi}%
. This is also consistent with the boundary state descriptions of O-planes.

For both cases of O-particle and O-domain-wall scatterings, one concluded that
no $s$-channel open string Regge poles existed since O-planes were known to be
not dynamical. Therefore, because the UV behavior of its scatterings is
power-law instead of exponential fall-off, the usual claim that there is a
thickness of order$\sqrt{\alpha^{^{\prime}}}$ for the O-domain-wall is misleading.

We summarize the pole structures of closed string states scattered from
various D-branes and O-planes in the following table. Since O-planes are not
dynamical, the $s$-channel open string Regge poles are not allowed.
Furthermore, for both cases of Domain-wall scatterings, the $t$-channel closed
string Regge poles are not allowed because there is only single kinematic
variable instead of two as in the usual cases.

\begin{center}
\ \
\begin{tabular}
[c]{|c|c|c|c|}\hline
& $p=-1$ & $1\leq p\leq23$ & $p=24$\\\hline
D$p$-branes & X & C+O & O\\\hline
O$p$-planes & X & C & X\\\hline
\end{tabular}

\end{center}

In this table, "C" and "O" represent infinite Closed string Regge poles and
Open string Regge poles respectively. "X" means that there are no infinite
Regge poles.

Following the suggestion of Mende \cite{Mende}, the authors In
\cite{Compact,Compact2} calculated high energy massive scattering amplitudes
of bosonic string with some coordinates compactified on the torus
\cite{Compact,Compact2}. Infinite linear relations among high energy
scattering amplitudes of different string states in the Hard scattering limit
were obtained. Furthermore, all possible power-law and exponential fall-off
regimes of high energy compactified bosonic string scatterings were analyzed
and classified by comparing the scatterings with their 26D noncompactified counterparts.

Interestingly, it was discovered in \cite{Compact2} that there exist a
power-law regime at fixed angle and an exponential fall-off regime at small
angle for high energy compactified open string scatterings \cite{Compact2}.
The linear relations break down as expected in all power-law regimes. The
analysis can be generalized to the high energy scatterings of the compactified
closed string, which corrects and extends the results in \cite{Compact}.

At this point, one may ask an important question for the results of
Eqs.(\ref{03}), (\ref{055}) and (\ref{04}) above , namely, is there any group
theoretical structure of the ratios of these scattering amplitudes? There is
indeed a simple analogy from the ratios of the nucleon-nucleon scattering
processes in particle physics,%
\begin{align}
(a)\text{ \ }p+p  &  \rightarrow d+\pi^{+},\nonumber\\
(b)\text{ \ }p+n  &  \rightarrow d+\pi^{0},\nonumber\\
(c)\text{ \ }n+n  &  \rightarrow d+\pi^{-}, \label{05}%
\end{align}
which can be calculated to be (ignore the mass difference between proton and
neutron)%
\begin{equation}
T_{a}:T_{b}:T_{c}=1:\frac{1}{\sqrt{2}}:1 \label{06}%
\end{equation}
from $SU(2)$ isospin symmetry. Is there any symmetry structure which can be
used to calculate ratios in Eqs.(\ref{03}), (\ref{055}), and (\ref{04})? It
turned out that part of the answer can be addressed by studying another high
energy regime of string scattering amplitudes, i.e. the fixed momentum
transfer or Regge regime (RR) which will be the main subject of the next
chapter \cite{RR1,RR2,RR3,RR4,RR5,RR6,RR7,bosonic,RRsusy,hep-th/0410131}.%

\setcounter{equation}{0}
\renewcommand{\theequation}{\arabic{section}.\arabic{equation}}%

\section{Stringy symmetries of Regge string scattering amplitudes}

In addition to the hard limit, i.e. the Gross regime (GR), which we have
reviewed in the previous chapters, another important regime is the high energy
limit with a small scattered angle, i.e. the Regge regime (RR). In this
chapter, we are going to review the Regge string scattering (RSS) amplitudes
and their relations to the HSS amplitudes. We will show that the number of RSS
amplitudes is much more numerous than that of HSS amplitudes. For example,
there are only $4$ HSS amplitudes as we discussed in Chapter 3, while $22$ RSS
amplitudes at mass level $M^{2}=4$ \cite{bosonic}. This is one of the reasons
why decoupling of ZNS in the RR, in contrast to the GR, is not good enough to
solve RSS amplitudes in terms of one single amplitude at each mass level.

The RR is defined as%
\begin{equation}
s\rightarrow\infty,\sqrt{-t}=\text{fixed (but }\sqrt{-t}\neq\infty).
\end{equation}
The relevant kinematics are%
\begin{subequations}
\begin{align}
e^{P}\cdot k_{1}  &  =-\frac{1}{M_{2}}\left(  \sqrt{p^{2}+M_{1}^{2}}%
\sqrt{p^{2}+M_{2}^{2}}+p^{2}\right)  \simeq-\frac{s}{2M_{2}},\\
e^{L}\cdot k_{1}  &  =-\frac{p}{M_{2}}\left(  \sqrt{p^{2}+M_{1}^{2}}%
+\sqrt{p^{2}+M_{2}^{2}}\right)  \simeq-\frac{s}{2M_{2}},\\
e^{T}\cdot k_{1}  &  =0 \label{111}%
\end{align}
and%
\end{subequations}
\begin{subequations}
\begin{align}
e^{P}\cdot k_{3}  &  =\frac{1}{M_{2}}\left(  \sqrt{q^{2}+M_{3}^{2}}\sqrt
{p^{2}+M_{2}^{2}}-pq\cos\theta\right)  \simeq-\frac{\tilde{t}}{2M_{2}}%
\equiv-\frac{t-M_{2}^{2}-M_{3}^{2}}{2M_{2}},\\
e^{L}\cdot k_{3}  &  =\frac{1}{M_{2}}\left(  p\sqrt{q^{2}+M_{3}^{2}}%
-q\sqrt{p^{2}+M_{2}^{2}}\cos\theta\right)  \simeq-\frac{\tilde{t}^{\prime}%
}{2M_{2}}\equiv-\frac{t+M_{2}^{2}-M_{3}^{2}}{2M_{2}},\\
e^{T}\cdot k_{3}  &  =-q\sin\phi\simeq-\sqrt{-{t}}.
\end{align}
Note that in contrast to the identification $e^{P}\simeq$ $e^{L}$ in the HSS
limit, we \textit{cannot} identify $e^{P}$ with $e^{L}$ in the RSS limit.
However, to compare with the HSS in the GR, we will consider the "relevant"
RSS amplitudes which contain polarizations $(e^{T},e^{L})$ only.

For illustration and to identify the ratios in Eqs.(\ref{03}) from RSS
amplitudes, we will first show an example at mass level $M^{2}=4$ in the RR.
In this case, there are eight high energy string amplitudes with polarizations
$(e^{T},e^{L})$ in the RR,%
\end{subequations}
\begin{align}
&  \alpha_{-1}^{T}\alpha_{-1}^{T}\alpha_{-1}^{T}|0\rangle,\alpha_{-1}%
^{L}\alpha_{-1}^{T}\alpha_{-1}^{T}|0\rangle,\alpha_{-1}^{L}\alpha_{-1}%
^{L}\alpha_{-1}^{T}|0\rangle,\alpha_{-1}^{L}\alpha_{-1}^{L}\alpha_{-1}%
^{L}|0\rangle,\nonumber\\
&  \alpha_{-1}^{T}\alpha_{-2}^{T}|0\rangle,\alpha_{-1}^{T}\alpha_{-2}%
^{L}|0\rangle,\alpha_{-1}^{L}\alpha_{-2}^{T}|0\rangle,\alpha_{-1}^{L}%
\alpha_{-2}^{L}|0\rangle.
\end{align}
Among them only four of the above amplitudes are relevant here and can be
calculated to be \cite{bosonic}%
\begin{align}
A^{TTT}  &  =\int_{0}^{1}dx\cdot x^{k_{1}\cdot k_{2}}\left(  1-x\right)
^{k_{2}\cdot k_{3}}\cdot\left(  \frac{ie^{T}\cdot k_{1}}{x}-\frac{ie^{T}\cdot
k_{3}}{1-x}\right)  ^{3}\nonumber\\
&  \simeq-i\left(  \sqrt{-t}\right)  ^{3}\frac{\Gamma\left(  -\frac{s}%
{2}-1\right)  \Gamma\left(  -\frac{\tilde{t}}{2}-1\right)  }{\Gamma\left(
\frac{u}{2}+3\right)  }\cdot\left(  -\frac{1}{8}s^{3}+\frac{1}{2}s\right)  ,\\
& \nonumber\\
A^{LLT}  &  =\int_{0}^{1}dx\cdot x^{k_{1}\cdot k_{2}}\left(  1-x\right)
^{k_{2}\cdot k_{3}}\cdot\left(  \frac{ie^{T}\cdot k_{1}}{x}-\frac{ie^{T}\cdot
k_{3}}{1-x}\right)  \left(  \frac{ie^{L}\cdot k_{1}}{x}-\frac{ie^{L}\cdot
k_{3}}{1-x}\right)  ^{2}\nonumber\\
&  \simeq-i\left(  \sqrt{-t}\right)  \left(  -\frac{1}{2M_{2}}\right)
^{2}\frac{\Gamma\left(  -\frac{s}{2}-1\right)  \Gamma\left(  -\frac{\tilde{t}%
}{2}-1\right)  }{\Gamma\left(  \frac{u}{2}+3\right)  }\nonumber\\
&  \cdot\left[  \left(  \frac{1}{4}t-\frac{9}{2}\right)  s^{3}+\left(
\frac{1}{4}t^{2}+\frac{7}{2}t\right)  s^{2}+\frac{\left(  t+6\right)  ^{2}}%
{2}s\right]  ,\\
& \nonumber\\
A^{TL}  &  =\int_{0}^{1}dx\cdot x^{k_{1}\cdot k_{2}}\left(  1-x\right)
^{k_{2}\cdot k_{3}}\cdot\left(  \frac{ie^{T}\cdot k_{1}}{x}-\frac{ie^{T}\cdot
k_{3}}{1-x}\right)  \left[  \frac{e^{L}\cdot k_{1}}{x^{2}}+\frac{e^{L}\cdot
k_{3}}{\left(  1-x\right)  ^{2}}\right] \nonumber\\
&  \simeq i\left(  \sqrt{-t}\right)  \left(  -\frac{1}{2M_{2}}\right)
\frac{\Gamma\left(  -\frac{s}{2}-1\right)  \Gamma\left(  -\frac{\tilde{t}}%
{2}-1\right)  }{\Gamma\left(  \frac{u}{2}+3\right)  }\nonumber\\
&  \left[  -\left(  \frac{1}{8}t+\frac{3}{4}\right)  s^{3}-\frac{1}{8}\left(
t^{2}-2t\right)  s^{2}-\left(  \frac{1}{4}t^{2}-t-3\right)  s\right]  ,
\end{align}
and%
\begin{align}
A^{LT}  &  =\int_{0}^{1}dx\cdot x^{k_{1}\cdot k_{2}}\left(  1-x\right)
^{k_{2}\cdot k_{3}}\cdot\left(  \frac{ie^{L}\cdot k_{1}}{x}-\frac{ie^{L}\cdot
k_{3}}{1-x}\right)  \left[  \frac{e^{T}\cdot k_{1}}{x^{2}}+\frac{e^{T}\cdot
k_{3}}{\left(  1-x\right)  ^{2}}\right] \nonumber\\
&  \simeq i\left(  \sqrt{-t}\right)  \left(  -\frac{1}{2M_{2}}\right)
\frac{\Gamma\left(  -\frac{s}{2}-1\right)  \Gamma\left(  -\frac{\tilde{t}}%
{2}-1\right)  }{\Gamma\left(  \frac{u}{2}+3\right)  }\cdot\left[  \frac{3}%
{4}s^{3}-\frac{t}{4}s^{2}-\left(  \frac{t}{2}+3\right)  s\right]  .
\end{align}
where the kinematic variables $(s,t)$ were used for convenient instead of
$(E,\theta)$ used in the GR. The conversion between the kinematic variables is
straightforward. From the above expressions, it is easy to show that all the
string amplitudes in the RR have the same leading order $\left(  \sim
s^{3}\right)  $ in energy. On the other hand, we note that, some terms. e.g.
$\sqrt{-t}t^{2}s^{2}$ in $A^{LLT}$ and $A^{TL}$, are of the leading order in
the GR, but subleading order in the RR; while some other terms, e.g.
$\sqrt{-t}s^{3}$ in $A^{LLT}$ and $A^{TL}$, are of the subleading order in the
GR, but leading order in the RR. This implies that the high energy string
scattering amplitudes in the GR and RR are complementary to each other.

To compare with the string amplitudes in the GR, we consider the high energy
string amplitudes in the RR with the same structure as those in the GR in
Eq.(\ref{Nmq}). The amplitudes $A^{TTT}$, $A^{LLT}$ , $A^{TL}$ and $A^{LT}$ at
mass level $M^{2}=4$ are the examples. For these string amplitudes, the
coefficients of the highest power of $t$ in the leading order amplitudes of
the RR are proportional to \cite{bosonic}
\begin{align}
A^{TTT}  &  =-i\left(  \sqrt{-t}\right)  \frac{\Gamma\left(  -\frac{s}%
{2}-1\right)  \Gamma\left(  -\frac{\tilde{t}}{2}-1\right)  }{\Gamma\left(
\frac{u}{2}+3\right)  }\cdot\left(  \frac{1}{8}ts^{3}\right)  \sim\frac{1}%
{8},\\
A^{LLT}  &  =-i\left(  \sqrt{-t}\right)  \left(  -\frac{1}{2M_{2}}\right)
^{2}\frac{\Gamma\left(  -\frac{s}{2}-1\right)  \Gamma\left(  -\frac{\tilde{t}%
}{2}-1\right)  }{\Gamma\left(  \frac{u}{2}+3\right)  }\left(  \frac{1}%
{4}ts^{3}\right)  \sim\frac{1}{64},\\
A^{TL}  &  =i\left(  \sqrt{-t}\right)  \left(  -\frac{1}{2M_{2}}\right)
\frac{\Gamma\left(  -\frac{s}{2}-1\right)  \Gamma\left(  -\frac{\tilde{t}}%
{2}-1\right)  }{\Gamma\left(  \frac{u}{2}+3\right)  }\cdot\left(  -\frac{1}%
{8}ts^{3}\right)  \sim-\frac{1}{32},\\
A^{LT}  &  =i\left(  \sqrt{-t}\right)  \left(  -\frac{1}{2M_{2}}\right)
\frac{\Gamma\left(  -\frac{s}{2}-1\right)  \Gamma\left(  -\frac{\tilde{t}}%
{2}-1\right)  }{\Gamma\left(  \frac{u}{2}+3\right)  }\cdot\left(  \frac{3}%
{4}\frac{1}{t}ts^{3}\right)  \sim0,
\end{align}
which produces the exactly same ratios in the GR in Eq.(\ref{03}). Here we
defined the symmetrized and anti-symmetrized amplitudes as%
\begin{align}
A^{\left(  TL\right)  }  &  =\frac{1}{2}\left(  A^{TL}+A^{LT}\right)
\sim\frac{1}{2}A^{TL},\\
A^{\left[  TL\right]  }  &  =\frac{1}{2}\left(  A^{TL}-A^{LT}\right)
\sim\frac{1}{2}A^{TL}.
\end{align}
It is interesting to see that $A^{LT}$ $\sim(\alpha_{-1}^{L})(\alpha_{-2}%
^{T})|0\rangle$ is in the subleading energy order in the GR, while it is in
the leading energy order in the RR, and it will not affect the ratios
calculated above.

From the above example of $M^{2}=4$, it was therefore believed that there
exist intimate connections between high energy string amplitudes in the GR and
RR. To study this link and to reproduce the ratios in Eq.(\ref{04}) in
particular, one was led to calculate RSS amplitudes at arbitrary mass levels.
Using the fact that $e^{T}\cdot k_{1}=0$ in Eq.(\ref{04}) and the energy power
counting, we obtain the following rules,%
\begin{align}
&  \alpha_{-n}^{T}:\quad\text{1 term (contraction of $ik_{3}\cdot X$ with
$\varepsilon_{T}\cdot\partial^{n}X$),}\\
&  \alpha_{-n}^{L}:%
\begin{cases}
n>1,\quad\text{1 term}\\
n=1\quad\text{2 terms}\text{ (contraction of $ik_{1}\cdot X$ and $ik_{3}\cdot
X$ with $\varepsilon_{L}\cdot\partial^{n}X$).}%
\end{cases}
\end{align}

The open string states with polarizations $(e^{T},e^{L})$ in the leading order
of the RR at each mass level $N=\sum_{n,l>0}np_{n}+lr_{l}$ are%
\begin{equation}
\left\vert p_{n},r_{l}\right\rangle =\prod_{n>0}(\alpha_{-n}^{T})^{p_{n}}%
\prod_{l>0}(\alpha_{-l}^{L})^{r_{l}}|0,k\rangle.
\end{equation}
The scattering string amplitudes of this state with three tachyonic states in
the $s-t$ channel can be calculated to be \cite{bosonic}
\begin{align}
A^{(p_{n},q_{m})}  &  =\left(  -\frac{i}{M_{2}}\right)  ^{q_{1}}U\left(
-q_{1},\frac{t}{2}+2-q_{1},\frac{\tilde{t}^{\prime}}{2}\right)  B\left(
-1-\frac{s}{2},-1-\frac{t}{2}\right) \nonumber\\
&  \cdot\prod_{n=1}\left[  i\sqrt{-t}(n-1)!\right]  ^{p_{n}}\prod_{m=2}\left[
i\tilde{t}^{\prime}(m-1)!\left(  -\frac{1}{2M_{2}}\right)  \right]  ^{q_{m}}.
\label{Apq}%
\end{align}
where $U(a,c,x)$ is the Kummer function of the second kind. It is crucial to
note that, in our formula, the parameter $c=t/2+2-q_{1}$ is not a constant
independent of the variable $x=\tilde{t}/2^{\prime}$, so that the function $U$
in the above amplitude does not satisfy the Kummer equation. On the other
hand, the parameter $a=-q_{1}$ is an integer, which causes that the Kummer
function in Eq.(\ref{Apq}) is truncated to a finite sum.

It can be seen from Eq.(\ref{Apq}) that the RSS amplitudes with spin
polarizations corresponding to Eq.(\ref{Nmq}) at each mass level are not
proportional to each other. Their ratios depend on $t$, so does the scattering
angle, and can be calculated to be \cite{bosonic}
\begin{align}
\frac{A^{(N,2m,q)}(s,t)}{A^{(N,0,0)}(s,t)}  &  =(-1)^{m}\left(  -\frac
{1}{2M_{2}}\right)  ^{2m+q}(\tilde{t}^{\prime}-2N)^{-m-q}(\tilde{t}^{\prime
})^{2m+q}\nonumber\\
&  \cdot\sum_{j=0}^{2m}(-2m)_{j}\left(  -1+N-\frac{\tilde{t}^{\prime}}%
{2}\right)  _{j}\frac{(-2/\tilde{t}^{\prime})^{j}}{j!}+\mathit{O}\left\{
\left(  \frac{1}{\tilde{t}^{\prime}}\right)  ^{m+1}\right\}  ,
\end{align}
where $(x)_{j}=x(x+1)(x+2)\cdots(x+j-1)$ is the Pochhammer symbol.

To reproduce the ratio in \ref{04}\/from the RSS for the general mass levels,
suggested by the explicit calculation at the mass level $M_{2}^{2}=4$
\cite{bosonic}, one needs to use the following identity,%
\begin{align}
&  \sum_{j=0}^{2m}(-2m)_{j}\left(  -L-\frac{\tilde{t}^{\prime}}{2}\right)
_{j}\frac{(-2/\tilde{t}^{\prime})^{j}}{j!}\nonumber\\
&  =0\cdot(-\tilde{t}^{\prime})^{0}\!+0\cdot(-\tilde{t}^{\prime})^{-1}%
\!+\dots+0\cdot(-\tilde{t}^{\prime})^{-m+1}\!+\frac{(2m)!}{m!}(-\tilde
{t}^{\prime})^{-m}+\mathit{O}\left\{  \left(  \frac{1}{\tilde{t}^{\prime}%
}\right)  ^{m+1}\right\}  \!
\end{align}
where $L=1-N$ is an integer. The identity was proved to be valid for any
non-negative integer $m$ and any \textit{real} number $L$ by using technique
of combinatorial number theory \cite{LYAM}. It was remarkable to first predict
\cite{bosonic} the mathematical identity above provided by string theory, and
then a rigorous mathematical proof followed \cite{LYAM}. It was also
interesting to see that the validity of the above identity includes
non-integer values of $L$ which were later shown to be realized by Regge
string scatterings in compact space \cite{HLY}. We thus have shown that the
ratios among HSS amplitudes calculated in Eqs.(\ref{03}) and (\ref{04}) can be
deduced and extracted from Kummer functions \cite{bosonic,KLY1,KLY2},
\begin{equation}
\frac{T^{(N,2m,q)}}{T^{(N,0,0)}}=\lim_{t\rightarrow\infty}\frac{A^{(N,2m,q)}%
}{A^{(N,0,0)}}=\left(  -\frac{1}{2M}\right)  ^{2m+q}2^{2m}\lim_{t\rightarrow
\infty}(-t)^{-m}U\left(  -2m\,,\,\frac{t}{2}+2-2m\,,\,\frac{t}{2}\right)  .
\label{07}%
\end{equation}
The above calculations can be generalized to the four classes of superstring
Regge scattering amplitudes \cite{RRsusy}.

The next interesting issue is to study relations among RSS amplitudes for
different string states. To achieve this, one considers the more general high
energy open string states with all three polarizations $(e^{T},e^{P},e^{L})$
in the RR at the mass level $N=\sum_{n,m,l>0}np_{n}+mq_{m}+lr_{l}$%
\begin{equation}
\left\vert p_{n},q_{m},r_{l}\right\rangle =\prod_{n>0}\left(  \alpha_{-n}%
^{T}\right)  ^{p_{n}}\prod_{m>0}\left(  \alpha_{-m}^{P}\right)  ^{q_{m}}%
\prod_{l>0}\left(  \alpha_{-l}^{L}\right)  ^{r_{l}}|0,k\rangle. \label{RRSS}%
\end{equation}
The string scattering amplitudes of the above state with three tachyonic
states in $s-t$ channel can be calculated to be%
\begin{align}
A^{(p_{n};q_{m};r_{l})}  &  =\int_{0}^{1}dx\,x^{k_{1}\cdot k_{2}}%
(1-x)^{k_{2}\cdot k_{3}}\cdot\left[  \frac{e^{P}\cdot k_{1}}{x}-\frac
{e^{P}\cdot k_{3}}{1-x}\right]  ^{q_{1}}\left[  \frac{e^{L}\cdot k_{1}}%
{x}+\frac{e^{L}\cdot k_{3}}{1-x}\right]  ^{r_{1}}\nonumber\\
&  \cdot\prod_{n=1}\left[  \frac{(n-1)!e^{T}\cdot k_{3}}{(1-x)^{n}}\right]
^{p_{n}}\prod_{m=2}\left[  \frac{(m-1)!e^{P}\cdot k_{3}}{(1-x)^{m}}\right]
^{q_{m}}\prod_{l=2}\left[  \frac{(l-1)!e^{L}\cdot k_{3}}{(1-x)^{l}}\right]
^{r_{l}}.
\end{align}
Finally, the string amplitudes can be expressed in two equivalent forms
\cite{LY}%
\begin{align}
A^{(p_{n};q_{m};r_{l})}  &  =\prod_{n>0}\left[  \left(  n-1\right)  !\sqrt
{-t}\right]  ^{p_{n}}\cdot\prod_{m>0}\left[  -\left(  m-1\right)
!\frac{\tilde{t}}{2M_{2}}\right]  ^{q_{m}}\cdot\prod_{l>1}\left[  \left(
l-1\right)  !\frac{\tilde{t}^{\prime}}{2M_{2}}\right]  ^{r_{l}}\nonumber\\
&  \quad\cdot B\left(  -\frac{s}{2}-1,-\frac{t}{2}+1\right)  \left(  \frac
{1}{M_{2}}\right)  ^{r_{1}}\nonumber\\
&  \cdot\sum_{i=0}^{q_{1}}\binom{q_{1}}{i}\left(  \frac{2}{\tilde{t}}\right)
^{i}\left(  -\frac{t}{2}-1\right)  _{i}U\left(  -r_{1},\frac{t}{2}%
+2-i-r_{1},\frac{\tilde{t}^{\prime}}{2}\right) \\
&  =\prod_{n>0}\left[  \left(  n-1\right)  !\sqrt{-t}\right]  ^{p_{n}}%
\cdot\prod_{m>1}\left[  -\left(  m-1\right)  !\frac{\tilde{t}}{2M}\right]
^{q_{m}}\cdot\prod_{l>0}\left[  \left(  l-1\right)  !\frac{\tilde{t}^{\prime}%
}{2M}\right]  ^{r_{l}}\nonumber\\
&  \cdot B\left(  -\frac{s}{2}-1,-\frac{t}{2}+1\right)  \left(  -\frac
{1}{M_{2}}\right)  ^{q_{1}}\nonumber\\
&  \cdot\sum_{j=0}^{r_{1}}\binom{r_{1}}{j}\left(  \frac{2}{\tilde{t}^{\prime}%
}\right)  ^{j}\left(  -\frac{t}{2}-1\right)  _{j}U\left(  -q_{1},\frac{t}%
{2}+2-j-q_{1},\frac{\tilde{t}}{2}\right)  . \label{kummer22}%
\end{align}
It is worth to note that, for $q_{1}=0$ or $r_{1}=0$, the RSS amplitudes can
be expressed in terms of a single Kummer function $U\left(  -r_{1},\frac{t}%
{2}+2-i-r_{1},\frac{\tilde{t}^{\prime}}{2}\right)  $ or $U\left(  -q_{1}%
,\frac{t}{2}+2-j-q_{1},\frac{\tilde{t}}{2}\right)  $. In general the RSS
amplitudes can be expressed in terms of a finite sum of Kummer functions, then
one can use the recurrence relations of Kummer functions to derive recurrence
relations among RSS amplitudes \cite{LY}.

For example, at mass level $M^{2}=4$, the recurrence relation%
\begin{equation}
U\left(  -3,\frac{t}{2}-1,\frac{t}{2}-1\right)  +\left(  \frac{t}{2}+1\right)
U(-2,\frac{t}{2}-1,\frac{t}{2}-1)-(\frac{t}{2}-1)U\left(  -2,\frac{t}{2}%
,\frac{t}{2}-1\right)  =0
\end{equation}
induces a recurrence relation among RSS amplitudes%
\begin{equation}
M\sqrt{-t}A^{PPP}-4A^{PPT}+M\sqrt{-t}A^{PPL}=0. \label{444}%
\end{equation}
Furthermore, the addition theorem of Kummer functions \cite{Slater}%
\begin{equation}
U(a,c,x+y)=\sum_{k=0}^{\infty}\frac{1}{k!}\left(  a\right)  _{k}(-1)^{k}%
y^{k}U(a+k,c+k,x)
\end{equation}
can be used to derive the inter-mass level recurrence relation of RSS
amplitudes. For example, by taking $a=-1,c=\frac{t}{2}+1,x=\frac{t}{2}-1$ and
$y=1,$ the theorem leads%
\begin{equation}
U\left(  -1,\frac{t}{2}+1,\frac{t}{2}\right)  -U\left(  -1,\frac{t}{2}%
+1,\frac{t}{2}-1\right)  -U\left(  0,\frac{t}{2}+2,\frac{t}{2}-1\right)  =0.
\end{equation}
which gives to an inter-mass level recurrence relation of RSS amplitudes
\cite{LY}%
\begin{equation}
M(2)(t+6)A_{2}^{TP}-2M(4)^{2}\sqrt{-t}A_{4}^{LP}+2M(4)A_{4}^{LT}=0 \label{rsr}%
\end{equation}
where $M(2)=\sqrt{2}$, $M(4)=\sqrt{4}=2,$ and $A_{2}$, $A_{4}$ are RSS
amplitudes at mass levels $M^{2}=2,4$. To derive Eq.(\ref{rsr}), it is crucial
to note that the power law behavior for each the RSS amplitude in
Eq.(\ref{rsr}) is independent on the mass level \cite{bosonic}.

Finally, Kummer recurrence relations can also be used to explicitly prove the
Regge stringy Ward identities obtained from decoupling of ZNS in the RR, but
not vice-versa. Thus in the RR, recurrence relations are more fundamental than
the linear relations obtained from decoupling of ZNS. However, we should keep
in mind that only Ward identities derived from the decoupling of Regge ZNS can
be generalized to the string loop amplitudes. As an example, it can be shown
that, in the Regge limit, the decoupling of the scalar type I Regge ZNS
\cite{LY}%
\begin{equation}
\lbrack25(\alpha_{-1}^{P})^{3}+9\alpha_{-1}^{P}(\alpha_{-1}^{L})^{2}%
+9\alpha_{-1}^{P}(\alpha_{-1}^{T})^{2}-9\alpha_{-2}^{L}\alpha_{-1}^{L}%
-9\alpha_{-2}^{T}\alpha_{-1}^{T}-75\alpha_{-2}^{P}\alpha_{-1}^{P}%
+50\alpha_{-3}^{P}]\left\vert 0,k\right\rangle \label{ward}%
\end{equation}
can be demonstrated by using the following recurrence relations of Kummer
functions%
\begin{align}
U(a-1,c,x)-(2a-c+x)U(a,c,x)+a(1+a-c)U(a+1,c,x)  &  =0,\\
U(a,c,x)-aU(a+1,c,x)-U(a,c-1,x)  &  =0,\\
\left(  c-a-1\right)  U(a,c-1,x)-\left(  x+c-1\right)  U\left(  a,c,x\right)
+xU\left(  a,c+1,x\right)   &  =0.
\end{align}

Similarly, infinite number of recurrence relations among RSS amplitudes at
arbitrary mass levels can be constructed. In general, these relations are
independent of stringy Ward identities derived from the decoupling of Regge ZNS.

However, in contrast to Ward identity obtained from the decoupling of Regge
ZNS like Eq.(\ref{ward}), we have no proof at loop levels for other ward
identities derived only from Kummer function recurrence relations. This is the
subtle difference between linear relations obtained in the GR and the
recurrence relations calculated in the RR. Similarly, one can construct
recurrence relations of higher spin generalization of the BPST vertex
operators \cite{RR6} by using the same way \cite{Tan}.

In general, each RSS amplitude is a sum of Kummer functions so that it becomes
complicated to derive the complete recurrence relations at higher mass levels.
In a later work \cite{AppellLY}, it was shown that the $26D$ open bosonic RSS
amplitude can be expressed in terms of a \textit{single} Appell function
$F_{1}$.

In fact, the $s-t$ channel RSS amplitudes with string state in Eq.(\ref{RRSS})
and three tachyons can be calculated as \cite{AppellLY}%
\begin{align}
A^{(p_{n};q_{m};r_{l})}  &  =\prod_{n=1}\left[  (n-1)!\sqrt{-t}\right]
^{p_{n}}\prod_{m=1}\left[  -(m-1)!\dfrac{\tilde{t}}{2M_{2}}\right]  ^{q_{m}%
}\prod_{l=1}\left[  (l-1)!\dfrac{\tilde{t}^{\prime}}{2M_{2}}\right]  ^{r_{l}%
}\nonumber\\
&  \cdot F_{1}\left(  -\tfrac{t}{2}-1,-q_{1},-r_{1},-\tfrac{s}{2};\dfrac
{s}{\tilde{t}},\dfrac{s}{\tilde{t}^{\prime}}\right)  \cdot B\left(  -\tfrac
{t}{2}-1,-\frac{s}{2}-1\right)  \label{abc}%
\end{align}
where the Appell function $F_{1}$ is one of the four generalizations of the
hypergeometric function $_{2}F_{1}$ to two variables, and is defined as%
\begin{equation}
F_{1}\left(  a;b,b^{\prime};c;x,y\right)  =\sum_{m=0}^{\infty}\sum
_{n=0}^{\infty}\dfrac{\left(  a\right)  _{m+n}\left(  b\right)  _{m}\left(
b^{\prime}\right)  _{n}}{m!n!\left(  c\right)  _{m+n}}x^{m}y^{n}%
\end{equation}
where $(a)_{n}=a\cdot\left(  a+1\right)  \cdots\left(  a+n-1\right)  $ is the
rising Pochhammer symbol. If $b$ or $b^{\prime}$ is a non-positive integer,
the Appell function would truncate to a finite polynomial that indeed is the
case for the Appell function in the RSS amplitudes obtained above. It should
be kept in mind that the expression in Eq.(\ref{abc}) is valid only when $s$
in the arguments of $F_{1}$ goes to $\infty$.

Using Appell function $F_{1}$, rather than a sum of Kummer functions, makes it
easier to obtain the complete recurrence relations among RSS amplitudes at
arbitrary mass levels, which have been conjectured to be associated to the
$SL(5,C)$ symmetry of $F_{1}$ \cite{sl5c}. For example, the recurrence
relation among RSS amplitudes \cite{AppellLY}%
\begin{equation}
\sqrt{-t}\left[  A^{(N;q_{1},r_{1})}+A^{(N;q_{1}-1,r_{1}+1)}\right]
-MA^{(N;q_{1}-1,r_{1})}=0 \label{555}%
\end{equation}
at arbitrary mass levels $M^{2}=2(N-1)$ can be obtained from recurrence
relations of the Appell functions. Eq.(\ref{555}) is a generalization of
Eq.(\ref{444}) to arbitrary mass levels. More general recurrence relations can
be obtained similarly. For example, from the leading term of $s$ in the Regge
limit, one obtain the following recurrence relation for $b_{2}$%
\begin{align}
cx^{2}F_{1}\left(  a;b_{1},b_{2};c;x,y\right)   & \nonumber\\
+\left[  \left(  a-b_{1}-b_{2}-1\right)  xy^{2}+cx^{2}-2cxy\right]
F_{1}\left(  a;b_{1},b_{2}+1;c;x,y\right)   & \nonumber\\
-\left[  \left(  a+1\right)  x^{2}y-\left(  a-b_{2}-1\right)  xy^{2}%
-cx^{2}+cxy\right]  F_{1}\left(  a;b_{1},b_{2}+2;c;x,y\right)   & \nonumber\\
-\left(  b_{2}+2\right)  x\left(  x-y\right)  yF_{1}\left(  a;b_{1}%
,b_{2}+3;c;x,y\right)   &  =0,
\end{align}
which induces to a recurrence relation for RSS amplitudes at arbitrary mass
levels \cite{AppellLY}%
\begin{align}
\tilde{t}^{\prime2}A^{(N;q_{1},r_{1})}  & \nonumber\\
+\left[  \tilde{t}^{\prime2}+\tilde{t}\left(  t-2\tilde{t}^{\prime}%
-2q_{1}-2r_{1}+4\right)  \right]  \left(  \frac{\frac{\tilde{t}^{\prime}}{2M}%
}{\sqrt{-t}}\right)  A^{(N;q_{1},r_{1}+1)}  & \nonumber\\
+\left[  \tilde{t}^{\prime2}-\tilde{t}^{\prime}\left(  \tilde{t}+t\right)
+\tilde{t}\left(  t-2r_{1}+4\right)  \right]  \left(  \frac{\frac{\tilde
{t}^{\prime}}{2M}}{\sqrt{-t}}\right)  ^{2}A^{(N;q_{1},r_{1}+2)}  & \nonumber\\
-2\left(  r_{1}-2\right)  \left(  \tilde{t}^{\prime}-\tilde{t}\right)  \left(
\frac{\frac{\tilde{t}^{\prime}}{2M}}{\sqrt{-t}}\right)  ^{3}A^{(N;q_{1}%
,r_{1}+3)}  &  =0.
\end{align}
More recurrence relations containing more than three Appell functions can be
found in \cite{Wang}.

More importantly, the recurrence relations of the Appell function $F_{1}$
\textit{in the Regge limit} can be systematically solved, so that one can
express all RSS amplitudes in terms of only one amplitude \cite{LY,AppellLY}.
All these results associate to symmetries of string scattering amplitudes in
hard limit discussed in chapter III
\cite{ChanLee,ChanLee1,ChanLee2,CHLTY1,CHLTY2,CHLTY3,susy}.

As the first step, we remind that there are two equivalent expressions
\cite{LY} as was previously shown in Eq.(\ref{kummer22}). It is easy to show
that, for $q_{1}=0$ or $r_{1}=0$, the RSS amplitudes can be expressed in terms
of a single Kummer function $U\left(  -r_{1},\frac{t}{2}+2-i-r_{1}%
,\frac{\tilde{t}^{\prime}}{2}\right)  $ or $U\left(  -q_{1},\frac{t}%
{2}+2-j-q_{1},\frac{\tilde{t}}{2}\right)  $, which are thus related to the
Appell function $F_{1}\left(  -\frac{t}{2}-1;0,-r_{1};\frac{s}{2};-\dfrac
{s}{\tilde{t}},-\dfrac{s}{\tilde{t}^{\prime}}\right)  $ or $F_{1}\left(
-\frac{t}{2}-1;-q_{1},0;\frac{s}{2};-\dfrac{s}{\tilde{t}},-\dfrac{s}{\tilde
{t}^{\prime}}\right)  $ respectively,%
\begin{align}
\lim_{s\rightarrow\infty}F_{1}\left(  -\frac{t}{2}-1;0,-r_{1};\frac{s}%
{2};-\dfrac{s}{\tilde{t}},-\dfrac{s}{\tilde{t}^{\prime}}\right)   &  =\left(
\frac{2}{\tilde{t}^{\prime}}\right)  ^{r_{1}}U\left(  -r_{1},\frac{t}%
{2}+2-r_{1},\frac{\tilde{t}^{\prime}}{2}\right)  ,\\
\lim_{s\rightarrow\infty}F_{1}\left(  -\frac{t}{2}-1;-q_{1},0;\frac{s}%
{2};-\dfrac{s}{\tilde{t}},-\dfrac{s}{\tilde{t}^{\prime}}\right)   &  =\left(
\frac{2}{\tilde{t}}\right)  ^{q_{1}}U\left(  -q_{1},\frac{t}{2}+2-q_{1}%
,\frac{\tilde{t}}{2}\right)  .
\end{align}
Furthermore, the ratio of Kummer functions can be obtained as \cite{LY},%
\begin{equation}
\frac{U(\alpha,\gamma,z)}{U(0,z,z)}=f(\alpha,\gamma,z),\alpha=0,-1,-2,-3,...
\end{equation}
where $f(\alpha,\gamma,z)$ can be obtained from the recurrence relations of
$U(\alpha,\gamma,z)$ and $U(0,z,z)=1$. In the Regge limit, one obtains that,
\begin{equation}
c=\dfrac{s}{2}\rightarrow\infty;x,y\rightarrow\infty;a,b_{1},b_{2}\text{
fixed,}%
\end{equation}
and the Appell functions $F_{1}\left(  a;0,b_{2};c;x,y\right)  $ and
$F_{1}\left(  a;b_{1},0;c;x,y\right)  $ are determined up to an overall factor
by recurrence relations. The next step is to obtain the recurrence relation%
\begin{equation}
yF_{1}\left(  a;b_{1},b_{2};c;x,y\right)  -xF_{1}\left(  a;b_{1}%
+1,b_{2}-1;c;x,y\right)  +\left(  x-y\right)  F_{1}\left(  a;b_{1}%
+1,b_{2};c;x,y\right)  =0, \label{1531}%
\end{equation}
which can be obtained from two of the four Appell recurrence relations among
contiguous functions.

We can now proceed to prove that in the Regge limit all RSS amplitudes can be
expressed in terms of a single amplitude. To be concise, we will use the
abbreviative notation $F_{1}\left(  a;b_{1},b_{2};c;x,y\right)  =F_{1}\left(
b_{1},b_{2}\right)  $ in the following argument. For $b_{2}=-1$, by using
Eq.(\ref{1531}) and the known $F_{1}\left(  b_{1},0\right)  $ and
$F_{1}\left(  0,b_{2}\right)  $, one can show that $F_{1}\left(
b_{1},-1\right)  $ can be determined for all $b_{1}=-1,-2,-3...$. Similarly,
$F_{1}\left(  b_{1},-2\right)  $ can be determined for all $b_{1}%
=-1,-2,-3...$.once $F_{1}\left(  b_{1},-1\right)  $ is known. This process can
be repeatedly used to determine $F_{1}\left(  b_{1},b_{2}\right)  $ for all
$b_{1},b_{2}=-1,-2,-3...$, so that all RSS amplitudes can be expressed in
terms of only one amplitude.%

\setcounter{equation}{0}
\renewcommand{\theequation}{\arabic{section}.\arabic{equation}}%

\section{The Lauricella string scattering amplitudes (LSSA)}

In this chapter, we briefly review the Lauricella string scattering amplitudes
(LSSA) discussed in a recent paper \cite{LLY2}. The authors considered the
four-point open bosonic string scattering amplitudes with three tachyons and
an arbitrary massive higher spin string state of the form,%
\begin{equation}
\left\vert r_{n}^{T},r_{m}^{P},r_{l}^{L}\right\rangle =\prod_{n>0}\left(
\alpha_{-n}^{T}\right)  ^{r_{n}^{T}}\prod_{m>0}\left(  \alpha_{-m}^{P}\right)
^{r_{m}^{P}}\prod_{l>0}\left(  \alpha_{-l}^{L}\right)  ^{r_{l}^{L}}%
|0,k\rangle, \label{state}%
\end{equation}
at the mass level $N=\sum_{n,m,l>0}\left(  nr_{n}^{T}+mr_{m}^{P}+lr_{l}%
^{L}\right)  $. The $\left(  s,t\right)  $ channel amplitude can be exactly
calculated and expressed in terms of the D-type Lauricella functions
\cite{LLY2}
\begin{align}
A_{st}^{(r_{n}^{T},r_{m}^{P},r_{l}^{L})}  &  =\prod_{n=1}\left[
-(n-1)!k_{3}^{T}\right]  ^{r_{n}^{T}}\cdot\prod_{m=1}\left[  -(m-1)!k_{3}%
^{P}\right]  ^{r_{m}^{P}}\prod_{l=1}\left[  -(l-1)!k_{3}^{L}\right]
^{r_{l}^{L}}\nonumber\\
&  \cdot B\left(  -\frac{t}{2}-1,-\frac{s}{2}-1\right)  F_{D}^{(K)}\left(
-\frac{t}{2}-1;R_{n}^{T},R_{m}^{P},R_{l}^{L};\frac{u}{2}+2-N;\tilde{Z}_{n}%
^{T},\tilde{Z}_{m}^{P},\tilde{Z}_{l}^{L}\right)  \label{stt}%
\end{align}
where%
\begin{align}
R_{k}^{X}  &  \equiv\left\{  -r_{1}^{X}\right\}  ^{1},\cdots,\left\{
-r_{k}^{X}\right\}  ^{k}\text{ with }\left\{  a\right\}  ^{n}%
=\underset{n}{\underbrace{a,a,\cdots,a}},\\
Z_{k}^{X}  &  \equiv\left[  z_{1}^{X}\right]  ,\cdots,\left[  z_{k}%
^{X}\right]  \text{ with }\left[  z_{k}^{X}\right]  =z_{k0}^{X},\cdots
,z_{k\left(  k-1\right)  }^{X}\text{,}\\
z_{kk^{\prime}}^{X}  &  =\left\vert \frac{k_{1}^{X}}{k_{3}^{X}}\right\vert
^{\frac{1}{k}}e^{\frac{2\pi ik^{\prime}}{k}}\text{ and }\tilde{z}_{kk^{\prime
}}^{X}\equiv1-z_{kk^{\prime}}^{X}\text{, }k^{\prime}=0,\cdots,k-1,
\end{align}
and the integer $K$ depends on the spin structure as%
\begin{equation}
\text{ }K=\underset{\{\text{for all }r_{j}^{T}\neq0\}}{\sum j}%
+\underset{\{\text{for all }r_{j}^{P}\neq0\}}{\sum j}+\underset{\{\text{for
all }r_{j}^{L}\neq0\}}{\sum j}.
\end{equation}
As a result, one can use the LSSA to reproduce SSA and the relations among SSA
of different string states at various scattering limits obtained previously.

In the HSS limit $e^{P}=e^{L}$ \cite{ChanLee1,ChanLee2}, $r_{1}^{T}=N-2m-2q$,
$r_{1}^{L}=2m$ and $r_{2}^{L}=q$, the LSSA can be calculated to be
\cite{LLY2}
\begin{align}
A_{st}^{(N-2m-2q,2m,q)}  &  \simeq B\left(  -\frac{t}{2}-1,-\frac{s}%
{2}-1\right)  \left(  E\sin\phi\right)  ^{N}\frac{\left(  2m\right)  !}%
{m!}\left(  -\frac{1}{2M_{2}}\right)  ^{2m+q}\nonumber\\
&  =(2m-1)!!\left(  -\frac{1}{M_{2}}\right)  ^{2m+q}\left(  \frac{1}%
{2}\right)  ^{m+q}A_{st}^{(N,0,0)},
\end{align}
which gives the ratios in Eq.(\ref{04}) that is the same as the previous
result \cite{ChanLee1,ChanLee2,CHL,CHLTY1,CHLTY2}.

In the RSS limit, the SSA in Eq.(\ref{stt}) reduces to \cite{AppellLY}%
\begin{align}
A_{st}^{(r_{n}^{T},r_{m}^{P},r_{l}^{L})}  &  \simeq B\left(  -\frac{t}%
{2}-1,-\frac{s}{2}-1\right)  \prod_{n=1}\left[  (n-1)!\sqrt{-t}\right]
^{r_{n}^{T}}\nonumber\\
&  \cdot\prod_{m=1}\left[  (m-1)!\frac{\tilde{t}}{2M_{2}}\right]  ^{r_{m}^{P}%
}\prod_{l=1}\left[  (l-1)!\frac{\tilde{t}^{\prime}}{2M_{2}}\right]
^{r_{l}^{L}}\nonumber\\
&  \cdot F_{1}\left(  -\frac{t}{2}-1;-q_{1},-r_{1};-\frac{s}{2};\frac
{s}{\tilde{t}},\frac{s}{\tilde{t}^{\prime}}\right)  , \label{appp}%
\end{align}
which agrees with the result obtained in \cite{AppellLY}.

Finally, in the NSS limit \cite{LLY1}, for the case of $r_{1}^{T}=N_{1}$,
$r_{1}^{P}=N_{3}$, $r_{1}^{L}=N_{2}$, and $r_{k}^{X}=0$ for all $k\geq2$, the
SSA reduces to%
\begin{align}
A_{st}^{(N_{1},N_{2},N_{3})}  &  =\left(  \frac{\epsilon}{2}\sin\phi\right)
^{N_{1}}\left(  \frac{\epsilon}{2}\cos\phi\right)  ^{N_{2}}\cdot\left(
-\frac{M_{1}+M_{2}}{2}\right)  ^{N_{3}}B\left(  \frac{M_{1}M_{2}}{2}%
,1-M_{1}M_{2}\right) \nonumber\\
&  \cdot\text{ }_{2}F_{1}\left(  \frac{M_{1}M_{2}}{2};-N_{3};M_{1}M_{2}%
;\frac{2M_{1}}{M_{1}+M_{2}}\right)  , \label{low}%
\end{align}
which agrees with the result obtained in \cite{LLY1}. The mass level dependent
of the string BCJ relation presented in Eq.(\ref{level}) can also be obtained
from Eq.(\ref{stt}).

Moreover, in \cite{LLLY} it was shown that by using the following key
recurrence relation of the Lauricella functions
\begin{equation}
x_{j}F_{D}^{(K)}\left(  b_{i}-1\right)  -x_{i}F_{D}^{(K)}\left(
b_{j}-1\right)  +\left(  x_{i}-x_{j}\right)  F_{D}^{(K)}=0,
\end{equation}
and a multiplication theorem of Gauss hypergeometric functions, one can
express all the LSSA in terms of the four tachyon amplitude. This result
extends the solvability of SSA at the HSS limit and the RSS limit discovered
previously to all kinematic regimes. We expect more interesting developments
on these research directions in the near future.

\section{Conclusion}

In addition to the string scatterings amplitudes at arbitrary mass level
discussed in this review, there were other related approaches in the
literature discussing higher spin dynamics of string theory. String theory
includes infinitely many higher spin massive fields with consistent mutual
interactions, and can provide useful hints on the dynamics of higher spin
field theory. On the other hand, a better understanding of higher spin
dynamics could also help our comprehension of string theory. It is widely
believed that the tensionless limit of string
\cite{Sagnotti,WS,WS1,WS2,WS3,WS4,WS5} is a theory of higher spin gauge
fields. An explicit and nontrivial construction of interacting higher spin
gauge theory is Vasiliev' system in AdS space-time.

In \cite{0305052}, the spectrum of Kaluza-Klein descendants of fundamental
string excitations on $AdS_{5}\times S^{5}$ was derived. Furthermore, in the
tensionless limit, the field equations from BRST quantization of string theory
provide a direct route toward local field equations for higher-spin gauge
fields \cite{0311257}.

Recently, in \cite{1207.4485}, one conjectured that the Vasiliev theory is a
limit of string theory. Roughly speaking, the fundamental string of string
theory is simply the flux tube string of the non-Abelian bulk Vasiliev theory.
The duality between Vasiliev theory and type IIA string field theory suggests
a concrete way of embedding Vasiliev theory into string theory. It is
interesting to investigate whether---and in what guise---the huge bulk gauge
symmetry of Vasiliev's description survives in the bulk string sigma model
description of the same system.

There existed other approaches of stringy symmetries which include other
studies of string collisions in the high energy, fixed momentum transfer
regime \cite{RR1,RR2,RR3,RR4,RR5,RR6,RR7}, the Hagedorn transition at high
temperature \cite{Hagedorn,Hagedorn1,hep-th/9908001}, vertex operator algebra
for compactified spacetime or on a lattice \cite{Moore,Moore1,CKT}, group
theoretical approach of string \cite{West1,West2}.

Another motivation of studying high energy string scattering is to investigate
the gravitational effect, such as black hole formation due to high energy
string collision, and to understand the nonlocal behavior of string theory.
Nevertheless, in \cite{0705.1816}, it was shown that there is no evidence that
the extendedness of strings produces any long-distance nonlocal effects in
high energy scattering, and no grounds have been found for string effects
interfering with formation of a black hole either.

\begin{acknowledgments}
We thank all of our collaborators to jointly work on this interesting subject.
We acknowledge financial supports from MoST, NCTS and ST Yau Center of NCTU.
\end{acknowledgments}

\end{document}